\documentclass[10pt]{iopart}
\usepackage{graphicx,epstopdf}
\usepackage{overpic}
\usepackage{subcaption}
\usepackage{esint}
\usepackage{xcolor}
\usepackage{hyperref}
\usepackage{amssymb,circle}
\definecolor{matlab_blue}{rgb}{0, 0.447, 0.741}
\definecolor{matlab_orange}{rgb}{0.85, 0.325, 0.098}
\definecolor{matlab_yellow}{rgb}{0.929, 0.694, 0.125}
\definecolor{matlab_green}{rgb}{0.466,0.674,0.188}
\definecolor{matlab_lightblue}{rgb}{0.301, 0.745, 0.933}
\definecolor{matlab_burgundy}{rgb}{0.635, 0.078, 0.184}
\definecolor{matlab_purple}{rgb}{0.494, 0.184, 0.556}
\definecolor{magenta}{rgb}{1, 0, 1}
\definecolor{blue}{rgb}{0, 0, 1}
\definecolor{red}{rgb}{1, 0, 0}
\definecolor{darkgreen}{rgb}{0.011, 0.128, 0.064}
\definecolor{black}{rgb}{0, 0, 0}
\definecolor{white}{rgb}{1, 1, 1}

\begin{document}

\title[]{Power exhaust and core-divertor compatibility of the baffled snowflake divertor in TCV}

\author{S Gorno$^1$, C Colandrea$^1$, O Février$^1$, H Reimerdes$^1$, C Theiler$^1$, B P Duval$^1$, T Lunt$^2$, H Raj$^3$, U A Sheikh$^1$, L Simons$^1$, A Thornton$^4$, the TCV Team\footnote{See author list of H. Reimerdes \textit{et al.} 2022 \textit{Nucl. Fusion} \textbf{62} 042018 (https://doi.org/10.1088/1741-4326/ac369b) for the TCV Team.} and the EUROfusion MST1 Team\footnote{See author list of B. Labit \textit{et al.} 2019 \textit{Nucl. Fusion} \textbf{59} 086020 (https://doi.org/10.1088/1741-4326/ab2211) for the EUROfusion MST1 Team.}}

\address{$^1$ Swiss Plasma Center (SPC), Ecole Polytechnique Fédérale de Lausanne (EPFL), Switzerland}
\address{$^2$ Max-Planck-Institut für Plasmaphysik, Garching bei München, Germany}
\address{$^3$ Institute for Plasma Research, Gandhinagar, India}
\address{$^4$ CCFE, Culham Science Centre, United Kingdom}
\ead{sophie.gorno@epfl.ch}
\vspace{10pt}
\begin{indented}
\item[]November 2022
\end{indented}

\begin{abstract} 

A baffled Snowflake Minus Low-Field Side (SF-LFS) is geometrically-optimised in TCV, increasing divertor neutral pressure, to evaluate the roles of divertor closure (comparing with an unbaffled SF-LFS) and magnetic geometry (comparing with a baffled Single Null, SN) in power exhaust and core-divertor compatibility. Ohmically-heated L-mode discharges in deuterium, with a line-averaged core density of approximately $4\times 10^{19}$ m$^{-3}$, are seeded with nitrogen to approach detached conditions. Baffles in the SF-LFS configuration are found to reduce the peak outer target heat flux by up to $23\%$, without significantly affecting the location of the inter-null radiation region or the core-divertor compatibility. When compared to the baffled SN, the baffled SF-LFS exhibits a reduction in outer target heat flux by up to $66\%$ and the ability to balance the strike-point distribution of heat flux. These benefits are less significant with N$_2$ seeding, with similar peak target quantities (such as heat flux, electron temperature and ion flux) and divertor radiated power. Despite a radiating region located farther from the confined plasma for the SF-LFS than the baffled SN, no change in core confinement is observed. Core effective charge even indicates an increase in core impurity penetration for the SF-LFS. These experiments constitute a good reference for detailed model validations and extrapolations, exploring important physics such as core impurity shielding and the dependence of divertor cross-field transport on magnetic geometry.

\end{abstract}

%
%
%
\ioptwocol

\section{Introduction}\label{sec:intro}

\hspace*{0.3cm}

In future fusion devices, material limits will constrain reactor operation, highlighting divertor detachment as an essential operational regime \cite{Pitts2019,Leonard2018}. Divertor vs main chamber neutral compression has been identified as a key parameter in accessing divertor detachment \cite{Leonard2018}: {future reactors will likely require operation at high divertor neutral pressures \cite{Pitts2019}. In TCV, regimes with enhanced divertor neutral pressure are achieved} by increasing divertor closure {with the installation of baffles }\cite{Fevrier2021,Reimerdes2021}. Alternative Divertor Configurations (ADCs) are being considered for future reactors to help mitigate heat fluxes arriving at the wall \cite{Reimerdes2020,Soukhanovskii2017,Theiler2017}. To this end, TCV's capacity to operate a wide range of ADCs has been enhanced by modifiable gas baffles \cite{fasoli2020}.

One such ADC is the Snowflake Minus Low-Field Side (SF-LFS) configuration \cite{Ryutov2015,Ryutov2007}, that features a secondary X-point in the LFS common flux region. The additional strike-points and extended region of low poloidal field in this geometry are modelled to increase divertor radiative losses and to lower target temperatures \cite{Reimerdes2013a,Pan2018,Giacomin2020c,Zhang2020}. A reduced peak outer target heat flux with respect to the standard Single Null (SN) configuration was already observed on TCV in the unbaffled SF-LFS, together with the ability to balance the ratio of power reaching each of the two active outer strike-points \cite{Maurizio2019e,Labit2017b}. 

Highly-radiative scenarios, achieved with impurity seeding, can greatly enhance divertor power dissipation. In the SN configuration, under detached conditions, an X-point radiator was observed with impurity seeding, where the dominant radiation is located in the vicinity of the X-point, reaching inside the confined region \cite{Bernert2017,Bernert2021,Fevrier2020}. Conversely, nitrogen seeded experiments in the unbaffled SF-LFS featured a strongly radiating region between the two X-points \cite{Reimerdes2017a}, that was further from the core plasma than for the SN. Reactor relevance will require the core impurities to remain below some critical concentration in order to maintain fusion performance and plasma stability \cite{Siccinio2019}. Therefore, assessing the impact of the radiation region displacement in the SF on core performance and core impurity levels is of high importance.

This study explores the performance of the baffled SF-LFS geometry, in terms of power exhaust and divertor-core compatibility, to appraise the combined effect of increased divertor closure and the complex magnetic geometry. In section \ref{sec:baffle_optimised}, the development of the baffled SF-LFS configuration is described. The divertor conditions of this geometry are then determined, beginning in section \ref{sec:target} with a comparison of the target heat fluxes with baffled SN and unbaffled SF-LFS geometries. Section \ref{sec:div_N2} explores the effect of N$_2$ seeding on these divertor conditions, assessing both the effect on divertor radiated power and mitigation of target heat fluxes. Section \ref{sec:core_conf} investigates the core-divertor compatibility of all configurations presented in this study, to compare the impact of magnetic geometries and divertor closures upon core impurity penetration and confinement. Finally, section \ref{sec:conc} presents the key conclusions and outlook of this study.

\section{Development of baffled SF-LFS configuration}\label{sec:baffle_optimised}

\hspace*{0.3cm}

\begin{figure*}[t]
	\centering
	\begin{subfigure}[t]{0.2\textwidth}
		\begin{overpic}[scale=0.65,percent]{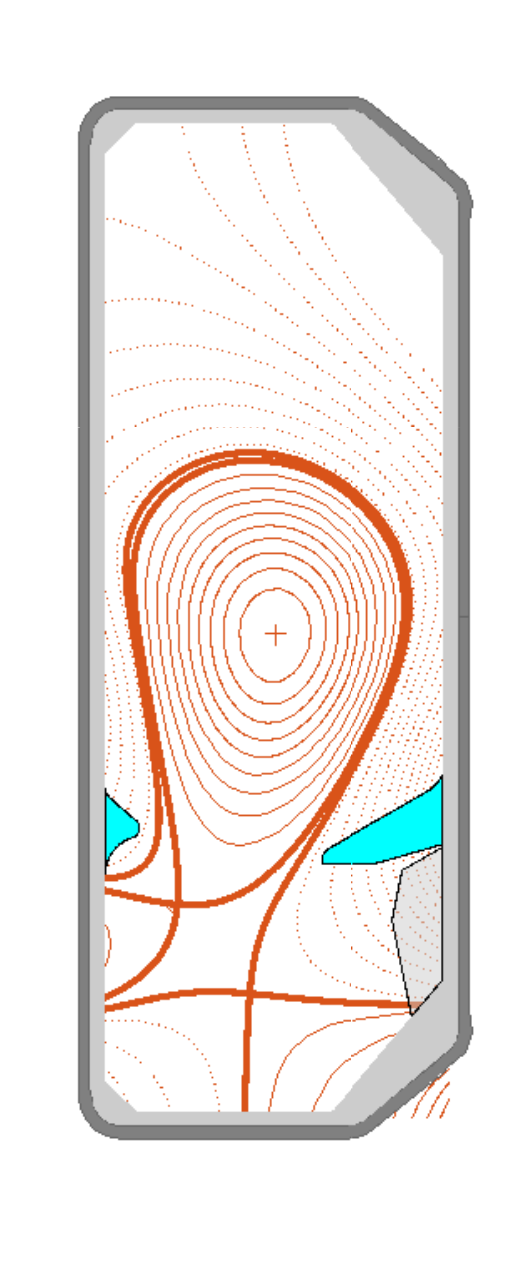}
			\put(9,100){\color{black}(a) SF-LFS}
			\put(9,95){\color{black}$dr_{\textrm{\scriptsize X2}}\sim6.9$ mm}
			\put(10,8){\color{matlab_orange}\tiny \#70497}
		\end{overpic}
	\end{subfigure}
	\begin{subfigure}[t]{0.2\textwidth}
		\begin{overpic}[scale=0.65,percent]{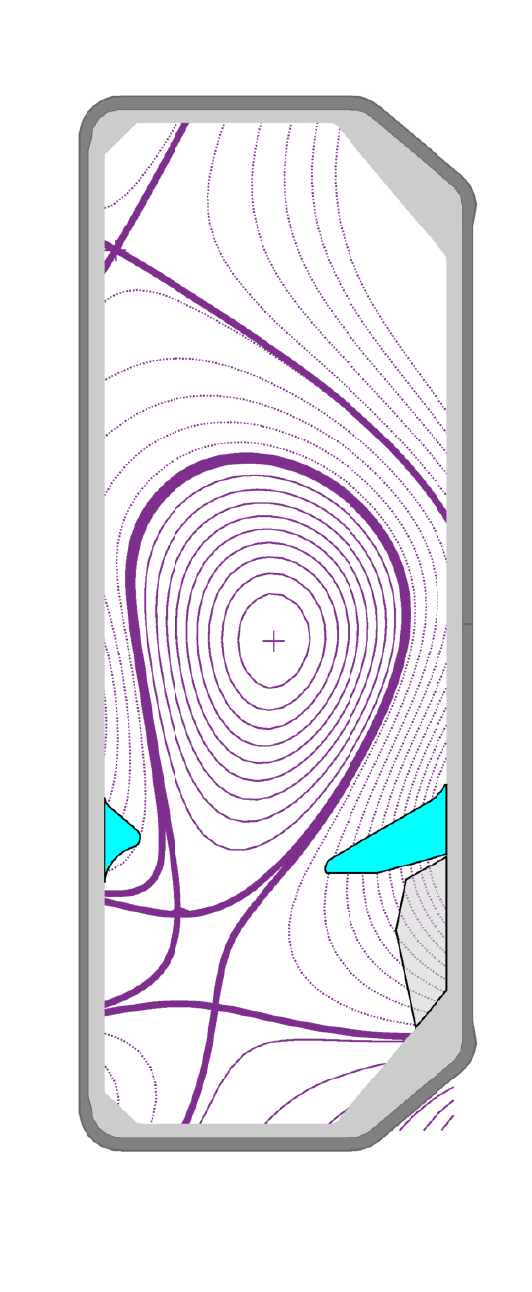}
			\put(9,100){\color{black}(b) SF-LFS}
			\put(9,95){\color{black}$dr_{\textrm{\scriptsize X2}}\sim3.7$ mm}
			\put(10,8){\color{matlab_purple}\tiny \#70498}
		\end{overpic}
	\end{subfigure}
	\begin{subfigure}[t]{0.2\textwidth}
		\begin{overpic}[scale=0.65,percent]{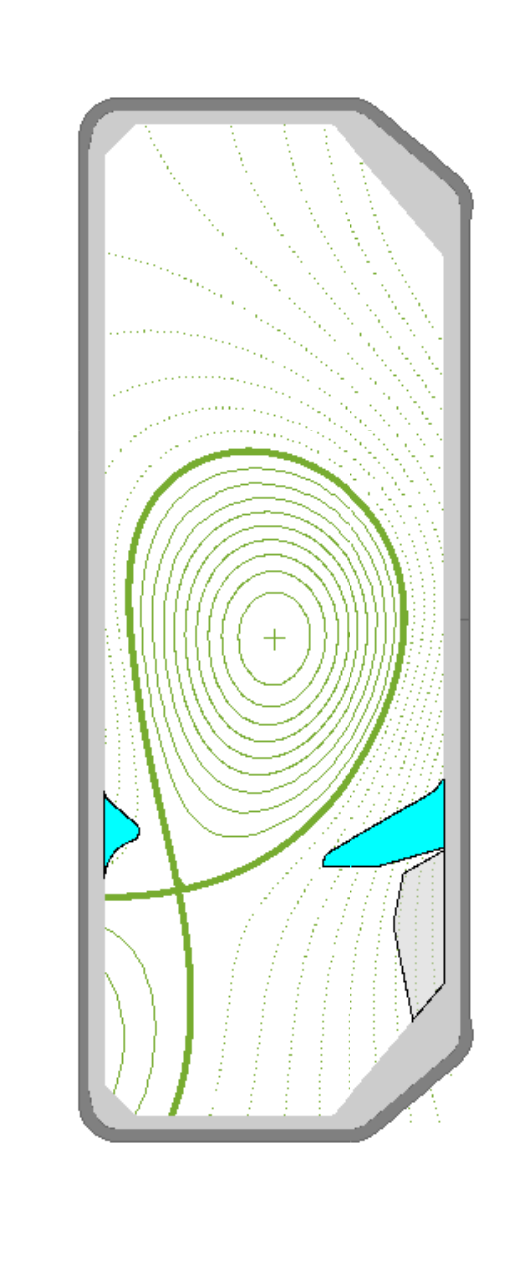}
			\put(9,100){\color{black}(c) SN}
			\put(10,8){\color{matlab_green}\tiny \#70322}
		\end{overpic}
	\end{subfigure}
	\begin{subfigure}[t]{0.3\textwidth}
		\begin{overpic}[scale=0.67,percent]{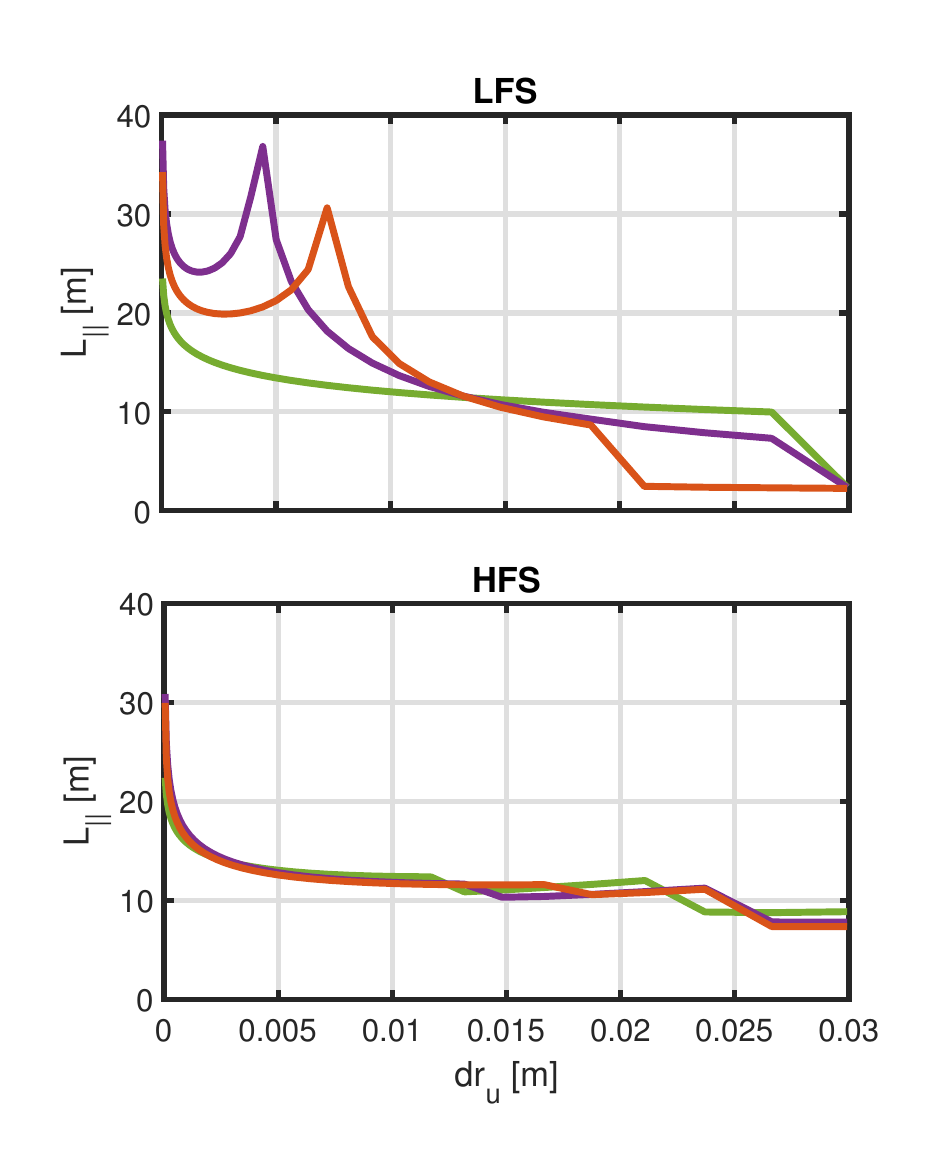}
			\put(65,85){\color{black}(d)}
			\put(65,43){\color{black}(e)}
		\end{overpic}
	\end{subfigure}
	\caption{(a)-(c) Poloidal view of the magnetic equilibrium reconstructions of baffle{-optimised} geometries. For each geometry, the parallel connection length $L_{||}$ is given from the outboard midplane to the (d) LFS and (e) HFS targets, as a function of the upstream distance from the primary separatrix, $dr_\textrm{\scriptsize u}$, with colours corresponding to the magnetic equilibria in (a)-(c).}
	\label{fig:geom_connectionlength}
\end{figure*}
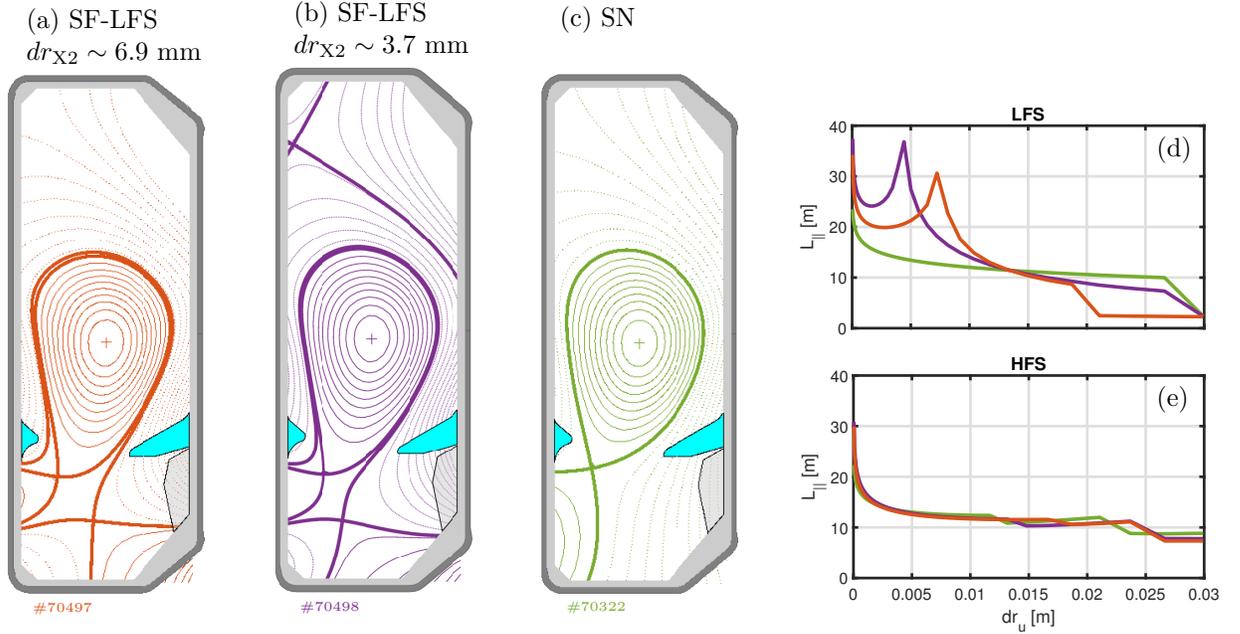

The first generation of TCV baffles were designed to maximise the core-to-divertor neutral compression in the standard SN geometry whilst remaining compatible with a variety of other divertor geometries \cite{fasoli2020} (see for example the study of long-legged ADCs with baffles \cite{Raj2022}). The SF-LFS features an even higher flux expansion in the null-point region than the SN, and so requires the development of a baffle-optimised geometry to maximise neutral compression whilst minimising plasma-baffle interaction. Two baffled SF-LFS geometries were developed with different X-point separations, together with a reference SN configuration, see figure \ref{fig:geom_connectionlength} (a)-(c). The SF-LFS configuration increases the LFS connection length with respect to the SN, with no compromise on the HFS connection length, figure \ref{fig:geom_connectionlength} (d)-(e). {Note that the plasma volume is up to 5\% lower in the SF-LFS compared to the SN.}

An experimental database of discharges was constructed for each of these three geometries, with and without baffles, to explore the effect of increased divertor closure. In this section, the experimental set-up of these discharges is first outlined, followed by a description of the divertor diagnostics used within the study. We then assess the effectiveness of this geometric baffle-optimisation, by comparing the baffled geometries with their unbaffled counterparts and initial, non-optimised baffled SF-LFS discharges, in terms of divertor neutral pressure and plasma-baffle interaction. 

\subsection{Experimental set-up}

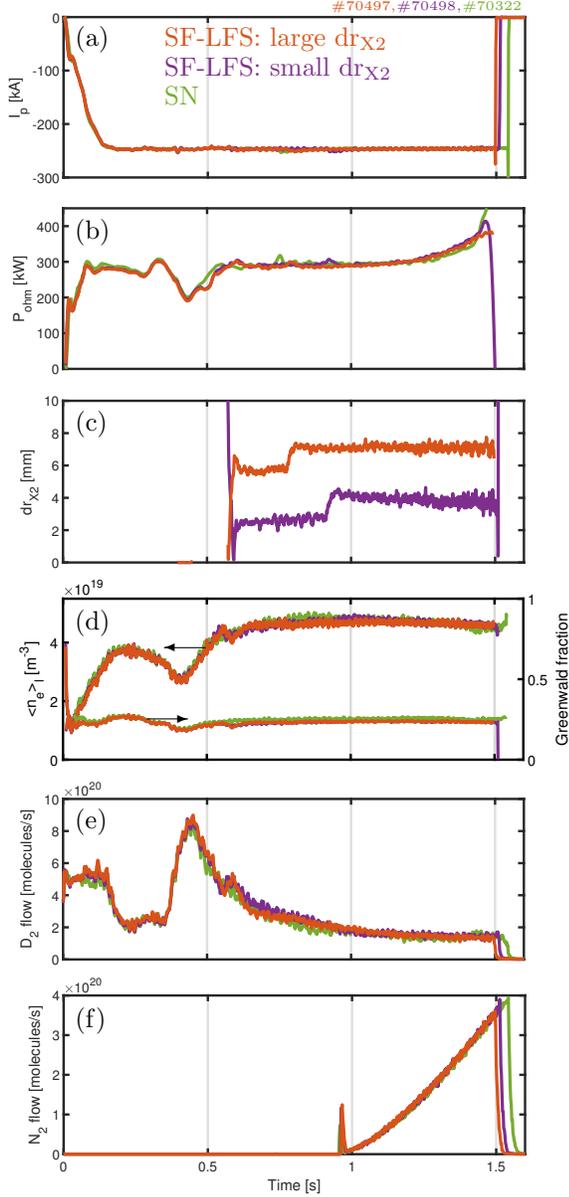
\begin{figure}[t]
	\centering
	\begin{subfigure}[r]{0.5\textwidth}
		\begin{overpic}[scale=0.6,percent]{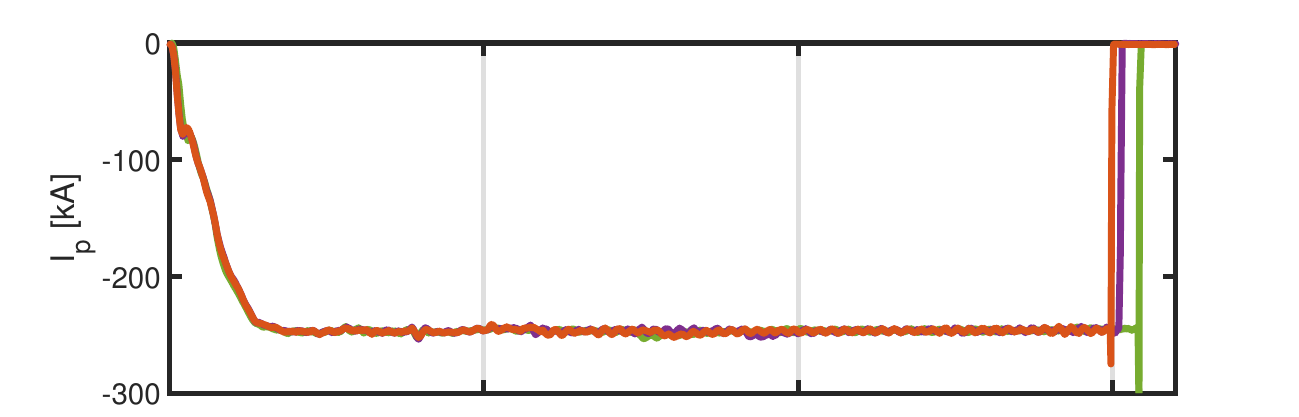}	
			\put(15,24){\color{black}(a)}
			\put(30,19){\color{matlab_purple}SF-LFS: small dr$_{\textrm{\scriptsize X2}}$}
			\put(30,24){\color{matlab_orange}SF-LFS: large dr$_{\textrm{\scriptsize X2}}$}
			\put(30,14){\color{matlab_green}SN}
			\put(58,30){\color{matlab_orange}\tiny \#70497,}
			\put(69,30){\color{matlab_purple}\tiny \#70498,}
			\put(80,30){\color{matlab_green}\tiny \#70322}
		\end{overpic}
	\end{subfigure}
	\begin{subfigure}[r]{0.5\textwidth}
		\begin{overpic}[scale=0.6,percent]{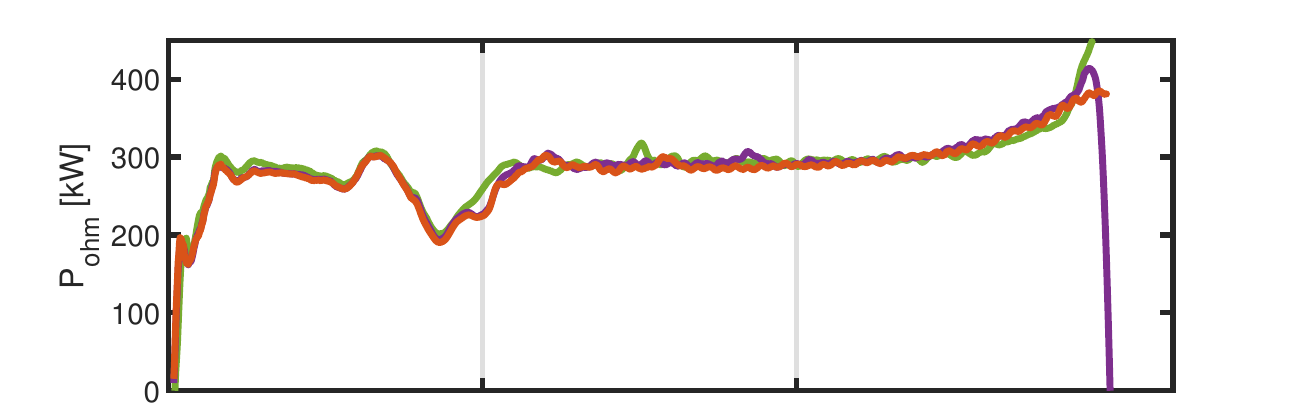}
			\put(15,24){\color{black}(b)}
		\end{overpic}
	\end{subfigure}
	\begin{subfigure}[r]{0.5\textwidth}
		\begin{overpic}[scale=0.6,percent]{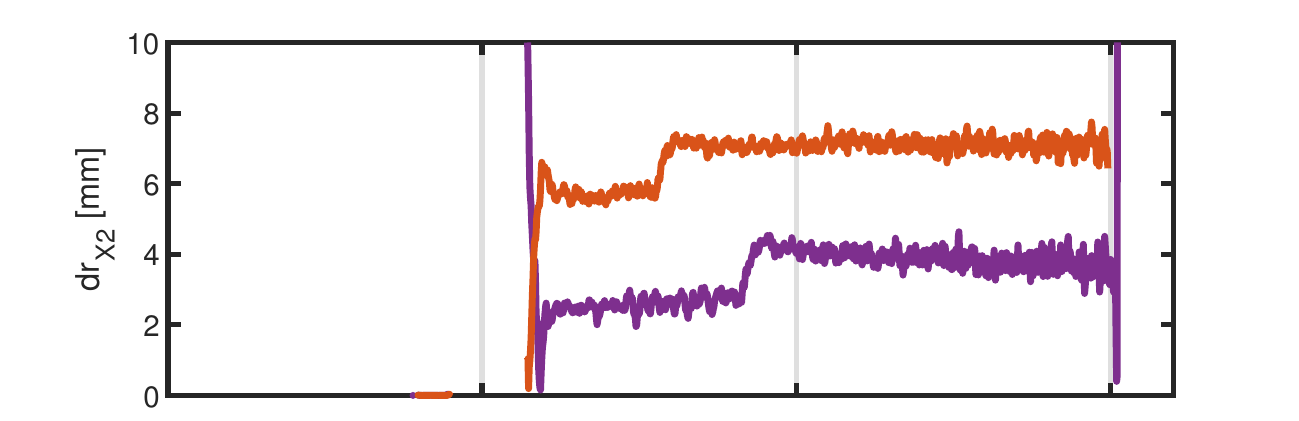}
			\put(15,24){\color{black}(c)}
		\end{overpic}
	\end{subfigure}
	\begin{subfigure}[r]{0.5\textwidth}
		\begin{overpic}[scale=0.6,percent]{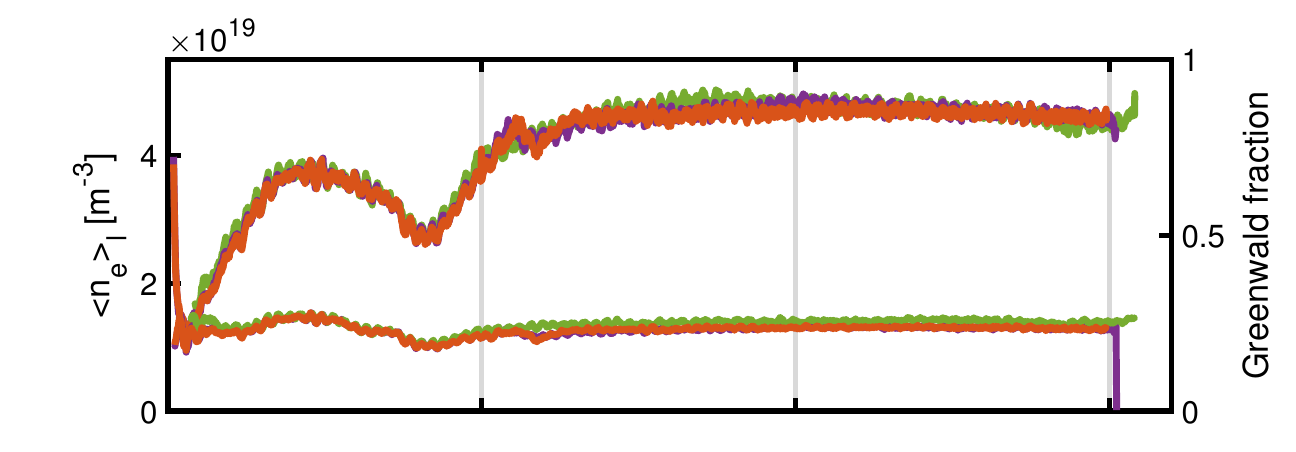}
			\put(15,25){\color{black}(d)}
			\put(37,22){\color{black}\vector(-2,0){7}}
			\put(27,10){\color{black}\vector(2,0){7}}
		\end{overpic}
	\end{subfigure}
	\begin{subfigure}[r]{0.5\textwidth}
		\begin{overpic}[scale=0.6,percent]{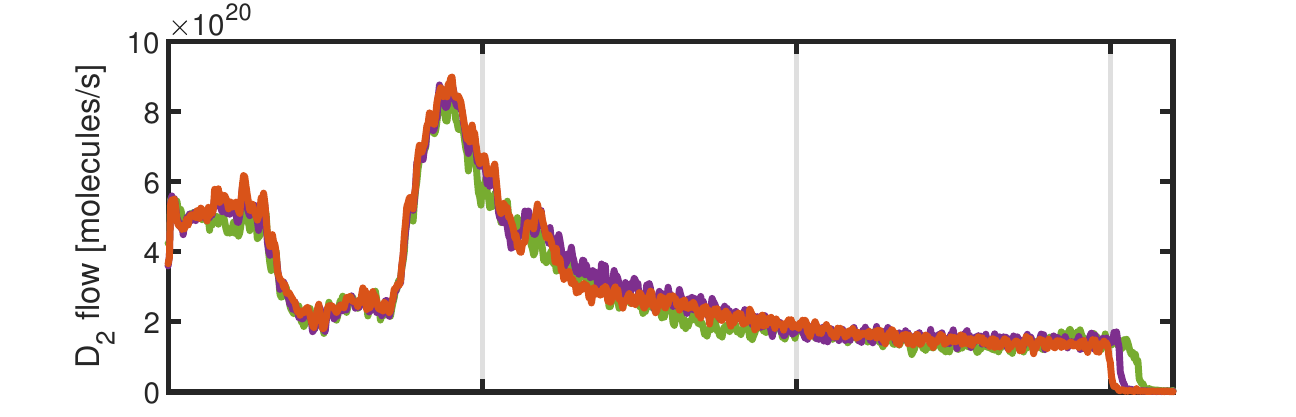}
			\put(15,24){\color{black}(e)}
		\end{overpic}
	\end{subfigure}
	\begin{subfigure}[r]{0.5\textwidth}
		\begin{overpic}[scale=0.6,percent]{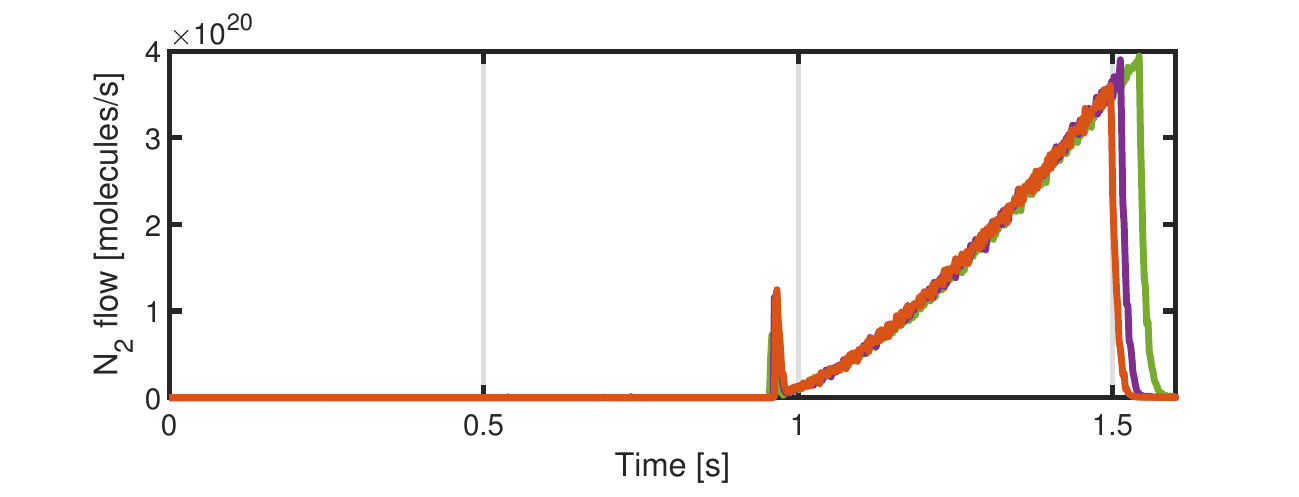}
			\put(15,29){\color{black}(f)}
		\end{overpic}
	\end{subfigure}
	\caption{Typical time traces of (a) $I_\textrm{\scriptsize p}$, (b) $P_{\textrm{\scriptsize ohm}}$, (c) $dr_{\textrm{\scriptsize X2}}$, (d) $\langle n_\textrm{\scriptsize e} \rangle_\textrm{\scriptsize l}$ and the Greenwald fraction, and the injected gas flow of (e) D$_2$ and (f) N$_2$ for each of the three baffled magnetic geometries: large $dr_{\textrm{\scriptsize X2}}$ SF-LFS (orange), small $dr_{\textrm{\scriptsize X2}}$ SF-LFS (purple), SN (green). The SF shape is established at $\sim~0.6$ s, with the desired X-point separation obtained at $\sim 0.80$ s and $\sim 0.96$ s for the large and small $dr_{\textrm{\scriptsize X2}}$ values, respectively.}
	\label{fig:timetraces}
\end{figure}

The SF-LFS and SN configurations are ohmically-heated L-mode discharges in deuterium, with a plasma current $I_p=245$ kA, operated in `reversed' toroidal magnetic field ($\nabla B$ ion drift directed upwards, away from the X-point) of $B_t=1.44$ T, figure \ref{fig:timetraces}. { This particular $B_t$ direction was chosen to avoid the H-mode transition and to facilitate detachment at the outer divertor.} The line-averaged core density is maintained at approximately $\langle n_e \rangle_l=4 .7\times 10^{19}$ m$^{-3}$, corresponding to a Greenwald fraction of $\sim0.25$, figure \ref{fig:timetraces} (d). N$_2$ is injected during the stationary phase of these discharges, to increase radiative cooling and approach detached conditions, figure \ref{fig:timetraces} (f). Several discharges were performed in each configuration, baffled and unbaffled, to increase diagnostic coverage. {With comparable core density and ohmic power traces in repeated discharges, we can compare these repeats for an indication of experimental reproducibility.}

In the SF-LFS geometry, the secondary X-point causes the outer scrape-off layer (SOL) to split into two outer strike-points (OSPs) magnetically-connected to the SOL, referred to as SP2 and SP4 \cite{Reimerdes2013a}. The X-point separation determines the power balance between SP2 and SP4 in attached conditions \cite{Maurizio2019e,Labit2017b}, and is quantified by $dr_{\textrm{\scriptsize X2}}$, the distance between the two separatrices at the outboard midplane. Accordingly, two SF-LFS geometries with different $dr_{\textrm{\scriptsize X2}}$ were developed to investigate the effect of divertor closure on the SF-LFS power sharing capability. The X-point separations in the baffled SF-LFS discharges are $dr_{\textrm{\scriptsize X2}}=6.9 \pm 0.2$ mm and $dr_{\textrm{\scriptsize X2}}=3.7 \pm 0.3$ mm (with errors corresponding to fluctuations within the discharge), figure \ref{fig:timetraces} (c). The heat flux decay width, $\lambda_q$, of the reference SN geometry is $\sim3.9$ mm \cite{Maurizio2018}: comparatively, these SF-LFS geometries represent `large' and `small' X-point separations respectively. Unbaffled discharges were also performed for the same X-point separations, with $dr_{\textrm{\scriptsize X2}}=6.6 \pm 0.7$ mm and $dr_{\textrm{\scriptsize X2}}=3.0 \pm 0.5$ mm.

\subsection{Divertor diagnostics} \label{subsec:Diags}

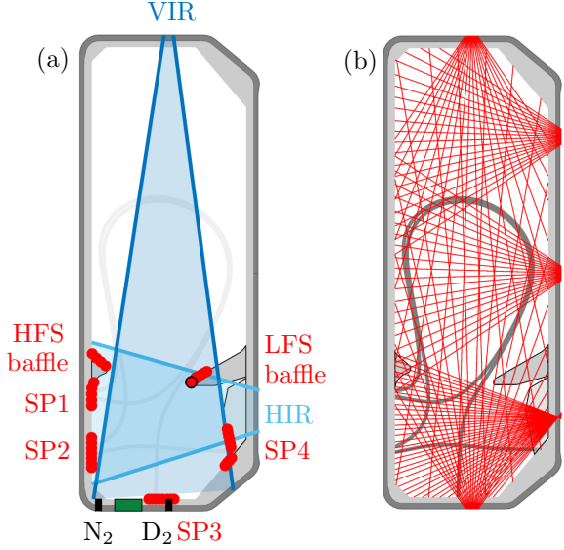
\begin{figure}[t]
	\centering
	\begin{subfigure}{0.23\textwidth}
	\centering
		\begin{overpic}[scale=0.7,percent]{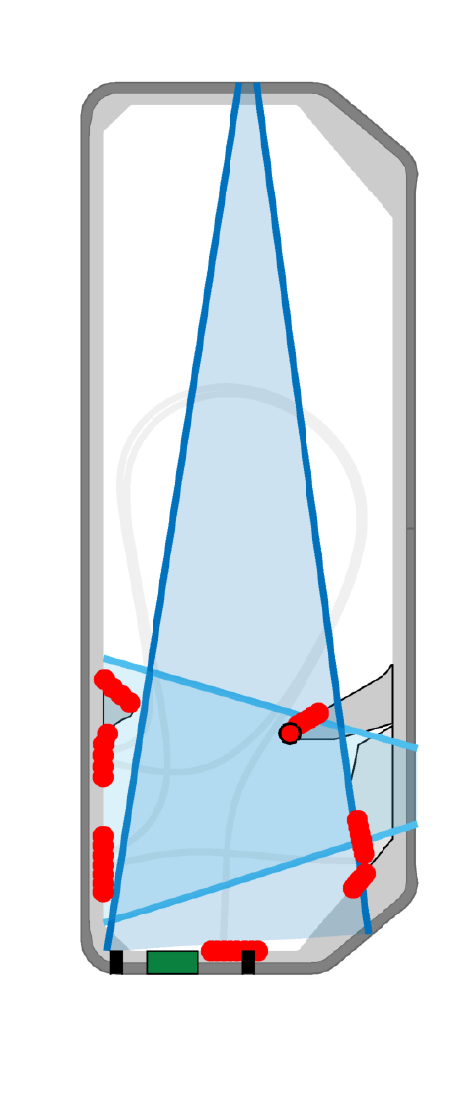}
			\put(0,87){\color{black}(a)}
			\put(-2,28){\color{red}SP1}
			\put(-2,20){\color{red}SP2}
			\put(24,6){\color{red}SP3}
			\put(39,20){\color{red}SP4}
			\put(-4,40){\color{red}HFS}
			\put(-5,35){\color{red}baffle}
			\put(39,38){\color{red}LFS}
			\put(39,33){\color{red}baffle}
			\put(8,6){\color{black}N$_2$}
			\put(18,6){\color{black}D$_2$}
			\put(39,26){\color{matlab_lightblue}HIR}
			\put(19,95){\color{matlab_blue}VIR}
		\end{overpic}
	\end{subfigure}
	\begin{subfigure}{0.23\textwidth}
		\centering
		\begin{overpic}[scale=0.7,percent]{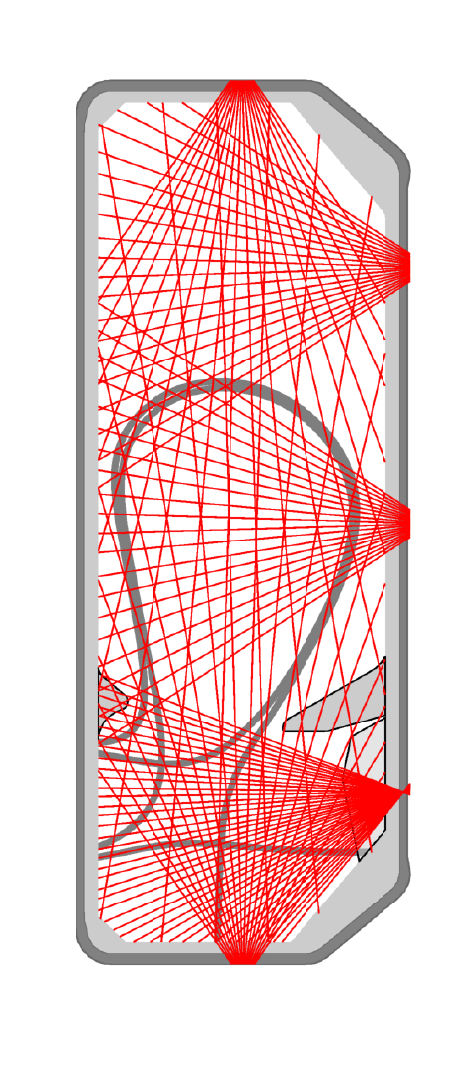}
			\put(0,87){\color{black}(b)}
		\end{overpic}
	\end{subfigure}
	\caption{Poloidal cross-section of the TCV vessel with the divertor diagnostics relevant to this study overplotted, (a):  Langmuir probes (red markers); vertical (VIR) and horizontal (HIR) infrared fields of view (blue areas); gas valves for D$_2$ and N$_2$ injection (black boxes); and pressure gauge (green box); (b): RADCAM bolometer lines of sight.}
	\label{fig:diags}
\end{figure}

Figure \ref{fig:diags} shows the poloidal position of the divertor diagnostics relevant to this study. The divertor neutral pressure is measured by a baratron pressure gauge at the divertor floor, shown in green in figure \ref{fig:diags} (a). The D$_2$ and N$_2$ injection valves, that include flow measurements, are located either side of this pressure gauge.

Langmuir probes cover all strike-points and the plasma-facing sides of the baffles, and are operated in swept bias mode to measure ion saturation current density, $j_{\textrm{\scriptsize sat}}$, target electron temperature, $T_e$, floating potential, $V_{\textrm{\scriptsize fl}}$, and the electric current when grounded, $j_0$, as detailed in \cite{Fevrier2018a,DeOliveira2019}. The target heat flux parallel to the magnetic field, $q_{||}$, can be calculated as the sum of contributions from electrons (e), ions (i) and the recombination process (rec), as performed in \cite{Brida2020},
\begin{eqnarray}\label{eq:q_gamma}
q_{||} = \underbrace{2T_e(j_{\textrm{\scriptsize sat}}-j_0)}_{q_{||,e}}+   \underbrace{(2.5T_e+eV_{sh})j_{\textrm{\scriptsize sat}}}_{q_{||,i}} + \underbrace{E_{\textrm{\scriptsize pot}}j_{\textrm{\scriptsize sat}}}_{q_{||,\textrm{\scriptsize rec}}},
\end{eqnarray}
where $E_{pot}=15.8$ eV is the potential energy per incident ion, including the recombination energy and half the molecular bonding energy, and $V_{sh}$ is the potential drop across the sheath,
\begin{equation}\label{eq:Vsh_gamma}
V_{\textrm{\scriptsize sh}}=-\frac{1}{2}\ln \left( \frac{4\pi m_e}{m_i} \right)T_e +V_{\textrm{\scriptsize fl}}.
\end{equation}
Here, we assume thermalisation between ions and electrons ($T_i=T_e$) and account for non-ambipolar conditions (non-zero SOL currents). 

The target heat flux is also measured by the vertical (VIR) and horizontal (HIR) infrared systems, using the temperature variation of the machine wall tiles \cite{Maurizio2018}. The fields of view of each system are portrayed in figure \ref{fig:diags} (a). The HIR measures heat fluxes at SP1 and SP2 for the SF-LFS and the SN's inner strike-point (ISP), whereas the VIR views the SN's OSP. Note that SP3 of the SF-LFS is inactive: it receives negligible heat and particle fluxes. IR measurements are used to complement the LP results, but remain incomplete as no IR data at SP4 is yet available. 

{In the following, we consider the target heat flux parallel to the magnetic field when comparing different divertor geometries, rather than that perpendicular to the target. The latter is strongly dependent on the incident magnetic field line angle, which is a function of both poloidal target flux expansion and wall tilt \cite{Theiler2017}. As the wall tilt is not optimised for the different configurations in this study, comparisons of the perpendicular heat flux could be misleading. }The {parallel} heat flux profiles from both LPs and IR are plotted in this work as a function of the normalised poloidal magnetic flux, $\rho_\psi=\sqrt{(\psi-\psi_0)/(\psi_{\textrm{\scriptsize LCFS}}-\psi_0)}$, where $\psi$ is the poloidal magnetic flux and $\psi_0$ and $\psi_{\textrm{\scriptsize LCFS}}$ are inferred at the magnetic axis and at the last closed flux surface (primary separatrix) respectively.

The plasma radiated power is measured by a recently-upgraded bolometry system, RADCAM \cite{Sheikh2022}, with coverage of both the core and divertor regions of TCV (see figure \ref{fig:diags} (b)). This system is used for the baffled discharges outlined above, but was not yet available for the older, unbaffled discharges. 

\subsection{Geometric baffle-optimisation} \label{sec:geom_baffle_optim}

An optimal distance between the SOL and LFS baffle exists where the divertor neutral pressure is maximised, due to competition between divertor closure and plasma-baffle interaction. Increasing further the SOL proximity to the LFS baffle increases the plasma recycling flux on the baffle and reduces divertor neutral pressure, weakening the benefit of gas baffles \cite{Galassi2020}. The SOL-baffle proximity is strongly affected by the plasma vertical position within the vessel, core plasma shape and $dr_{\textrm{\scriptsize X2}}$. Thus, baffle-optimised SF-LFS geometries were designed to maximise the SOL-baffle distance, while maintaining a sufficiently large $dr_{\textrm{\scriptsize X2}}$ to approach a balanced peak heat flux distribution at the OSPs. {This involved reducing the lower core width, hence decreasing plasma volume and increasing elongation, while shifting the X-points towards the HFS.}

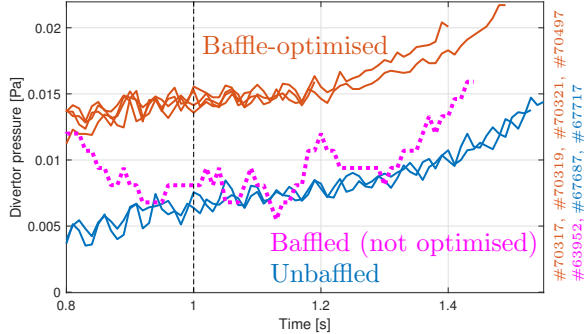
\begin{figure}[t]
	\centering
	\begin{overpic}[scale=0.5,percent]{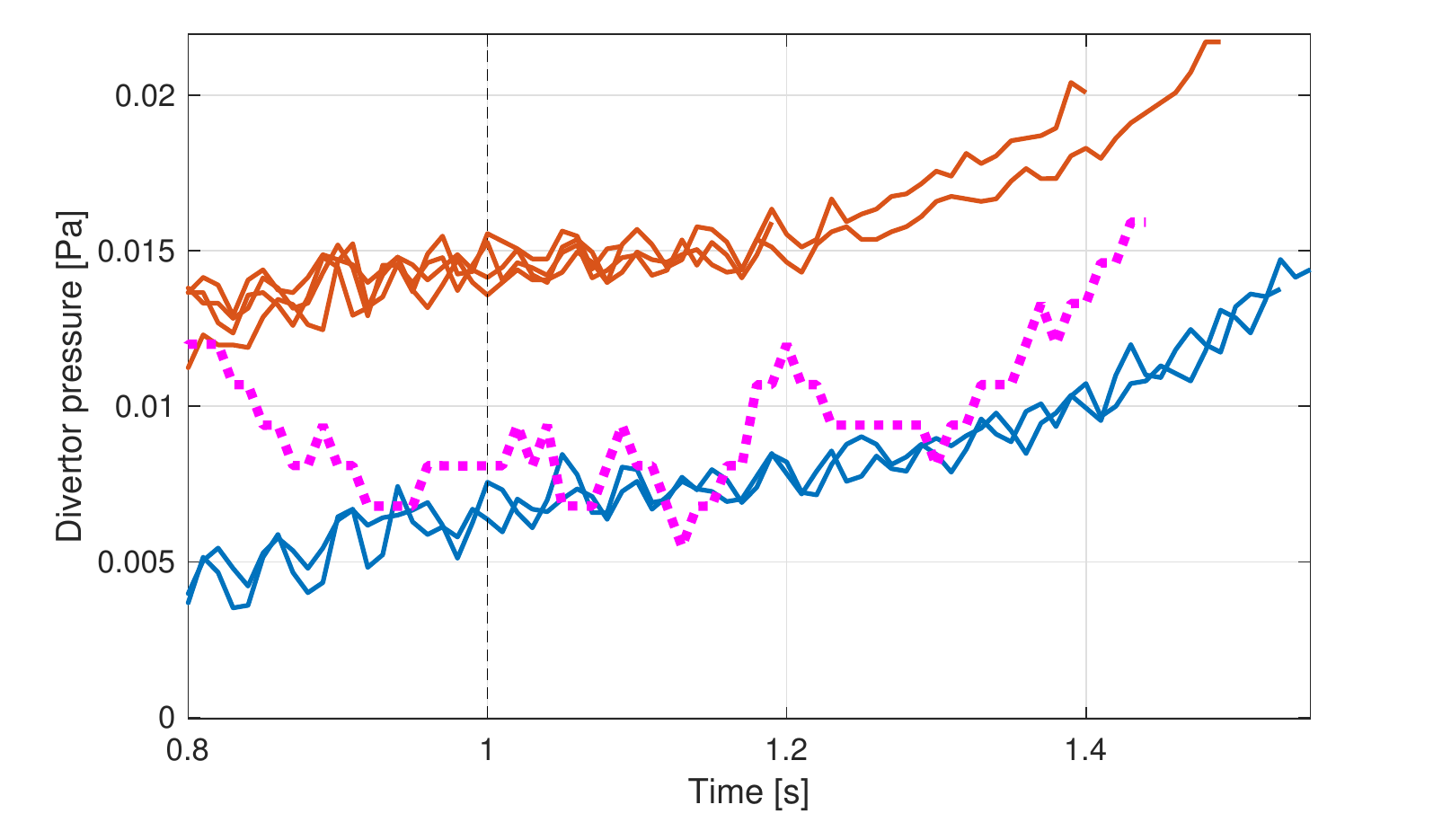}
		\put(35,47){\color{matlab_orange}Baffle-optimised}
		\put(46,10){\color{matlab_blue}Unbaffled}
		\put(46,15){\color{magenta}Baffled (not optimised)}
		\put(92,10){\rotatebox{90}{\color{matlab_orange}\tiny \#70317, \#70319, \#70321, \#70497}}
		\put(95,21){\rotatebox{90}{\color{matlab_blue}\tiny \#67687, \#67717}}
		\put(95,10){\rotatebox{90}{\color{magenta}\tiny \#63952,}}
	\end{overpic}
	\caption{Divertor neutral pressure measurements from the divertor floor pressure gauge for the unbaffled (blue), baffled but not optimised (magenta) and baffle-optimised (orange) SF-LFS configurations. The vertical dashed line at 1s represents the beginning of a N$_2$ seeding ramp. Results are shown from repeat discharges in each geometry.}
	\label{fig:divpress}
\end{figure}

For the baffle-optimised SF-LFS geometries, the divertor neutral pressure approximately doubles compared to the equivalent unbaffled geometry (shown in figure \ref{fig:divpress}), similarly to the baffled standard SN configuration at these upstream conditions \cite{Fevrier2021}. This difference is maintained throughout the discharges, even when seeding N$_2$ into the divertor. The first attempts of baffled SF-LFS discharges, that were not baffle-optimised, have comparable divertor neutral pressure to the unbaffled cases, demonstrating the need for attentive geometric optimisation. 

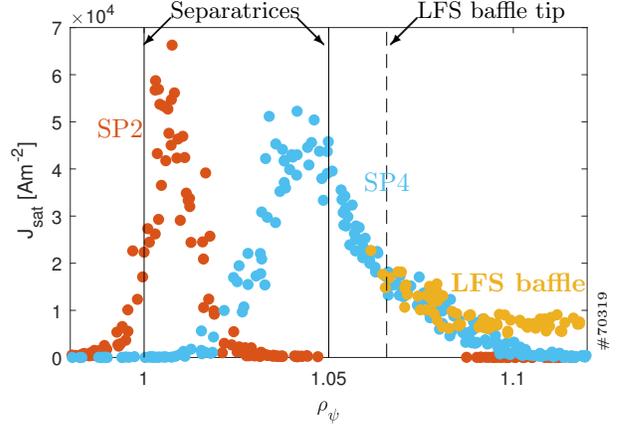
\begin{figure}[t]
	\centering
	\begin{overpic}[scale=0.7,percent]{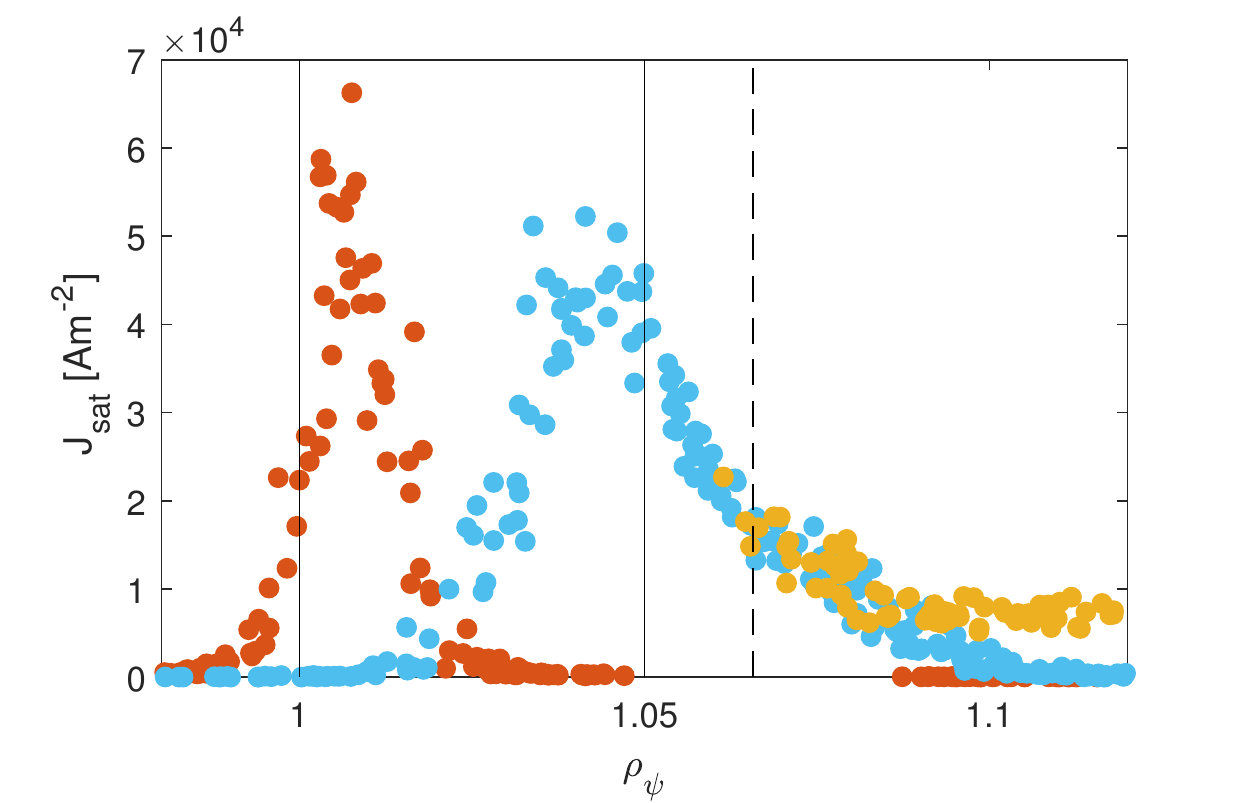}
		\put(17,43){\color{matlab_orange}SP2}
		\put(57,35){\color{matlab_lightblue}SP4}
		\put(70,20){\color{matlab_yellow}\bf{LFS baffle}}
		\put(92,11){\rotatebox{90}{\color{black}\tiny \#70319}}
		\put(28,61){\color{black}\small Separatrices}
		\put(28,61){\color{black}\vector(-1,-1){4}}
		\put(47,61){\color{black}\vector(1,-1){4}}
		\put(65,61){\color{black}\small LFS baffle tip}
		\put(65,61){\color{black}\vector(-1,-1){4}}
	\end{overpic}
	\caption{Profiles of the ion saturation current density parallel to the magnetic field ($j_{\textrm{\scriptsize sat}}$) measured by LPs at the LFS baffle and active outer strike-points (SP2, SP4) of the baffle-optimised SF-LFS configuration with $dr_{\textrm{\scriptsize X2}}\sim6.9$ mm. The separatrices are indicated by solid lines and the position of the LFS baffle tip is indicated by a dashed line, at $\rho_\psi=1.069\pm0.005$.}
	\label{fig:LPprofs_baffleinteraction}
\end{figure}

As expected, the baffle-optimisation also strongly reduces the particle flux impinging the plasma-facing side of the LFS baffle. In the baffle-optimised SF-LFS geometry, the magnetic flux surface $\rho_\psi \sim 1.07$ (corresponding to an upstream distance from the separatrix of $dr_u\sim22$ mm) is intercepted by the LFS baffle tip (see figure \ref{fig:LPprofs_baffleinteraction}). By constrast, for the geometry which is not baffle-optimised, the baffle tip intercepts $\rho_\psi \sim 1.04$ (correspondingly $dr_u\sim12$ mm), being much closer to the core plasma (not shown). The ion flux recycled at the plasma-facing LFS baffle surface is considerably reduced by the geometric optimisation, becoming comparable to that of the reference SN (in which the LFS baffle tip intercepts $\rho_\psi \sim 1.09$, $dr_u\sim27$ mm).

Note that the HFS baffle interaction remains negligible for all baffle-optimised geometries. The baffle-optimised SF-LFS configuration thus allows for a fairer comparison of target behaviour between configurations. 

\section{Target conditions of the baffled SF-LFS}\label{sec:target}

\hspace*{0.3cm}

In the SF-LFS geometry, the power distribution between the active outer strike-points depends upon the X-point separation: the higher $dr_{\textrm{\scriptsize X2}}$, the higher the ratio of the heat fluxes reaching SP2 to SP4. Of the two SF-LFS geometries considered in this study (see figure \ref{fig:geom_connectionlength}), the case with $dr_{\textrm{\scriptsize X2}}\sim3.7$ mm presents a more balanced peak heat flux ($q_{||}^{\textrm{\scriptsize peak}}$) distribution between the outer targets, while the case with $dr_{\textrm{\scriptsize X2}}\sim6.9$ mm presents a higher $q_{||}^{\textrm{\scriptsize peak}}$ at SP2 than SP4. 

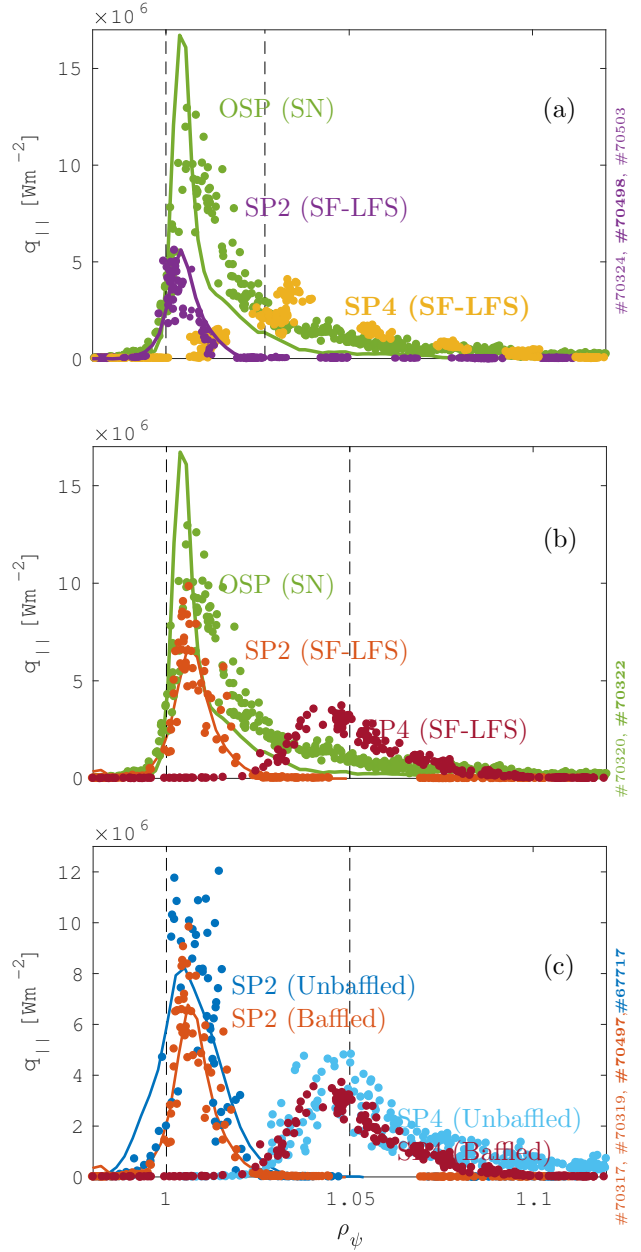
\begin{figure}[t]
	\centering
	\begin{subfigure}{0.5\textwidth}
		\centering
		\begin{overpic}[width=8.8cm]{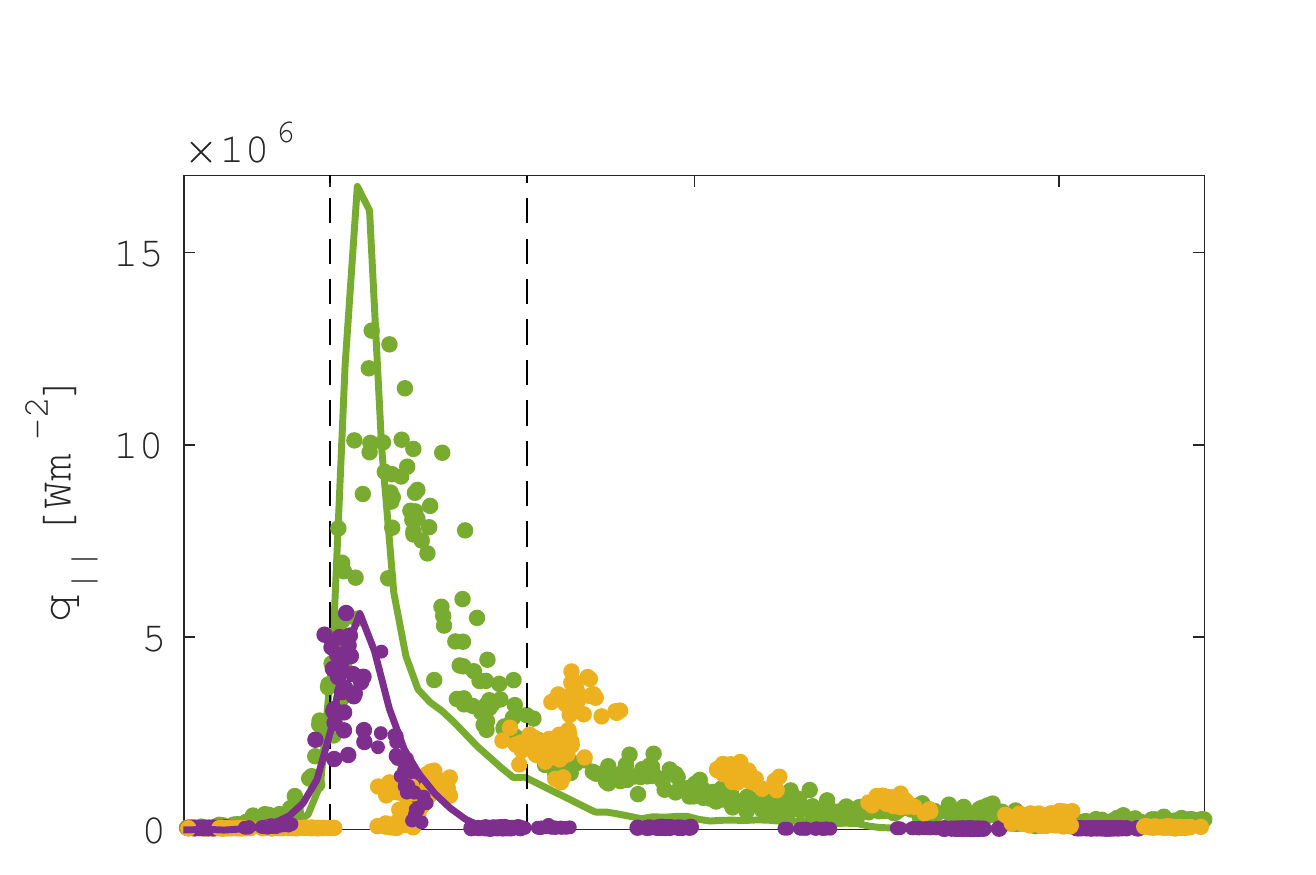}
			\put(37,25){\color{matlab_purple}SP2 (SF-LFS)}
			\put(52,10){\color{matlab_yellow}\bf{SP4 (SF-LFS)}}
			\put(33,40){\color{matlab_green}OSP (SN)}
			\put(82,40){\color{black}(a)}
			\put(93,10){\rotatebox{90}{\color{matlab_purple}\tiny \#70324, \textbf{\#70498}, \#70503}}
		\end{overpic}
	\end{subfigure}
	\begin{subfigure}{0.5\textwidth}
		\centering
		\vspace{0.05cm}
		\begin{overpic}[width=8.8cm]{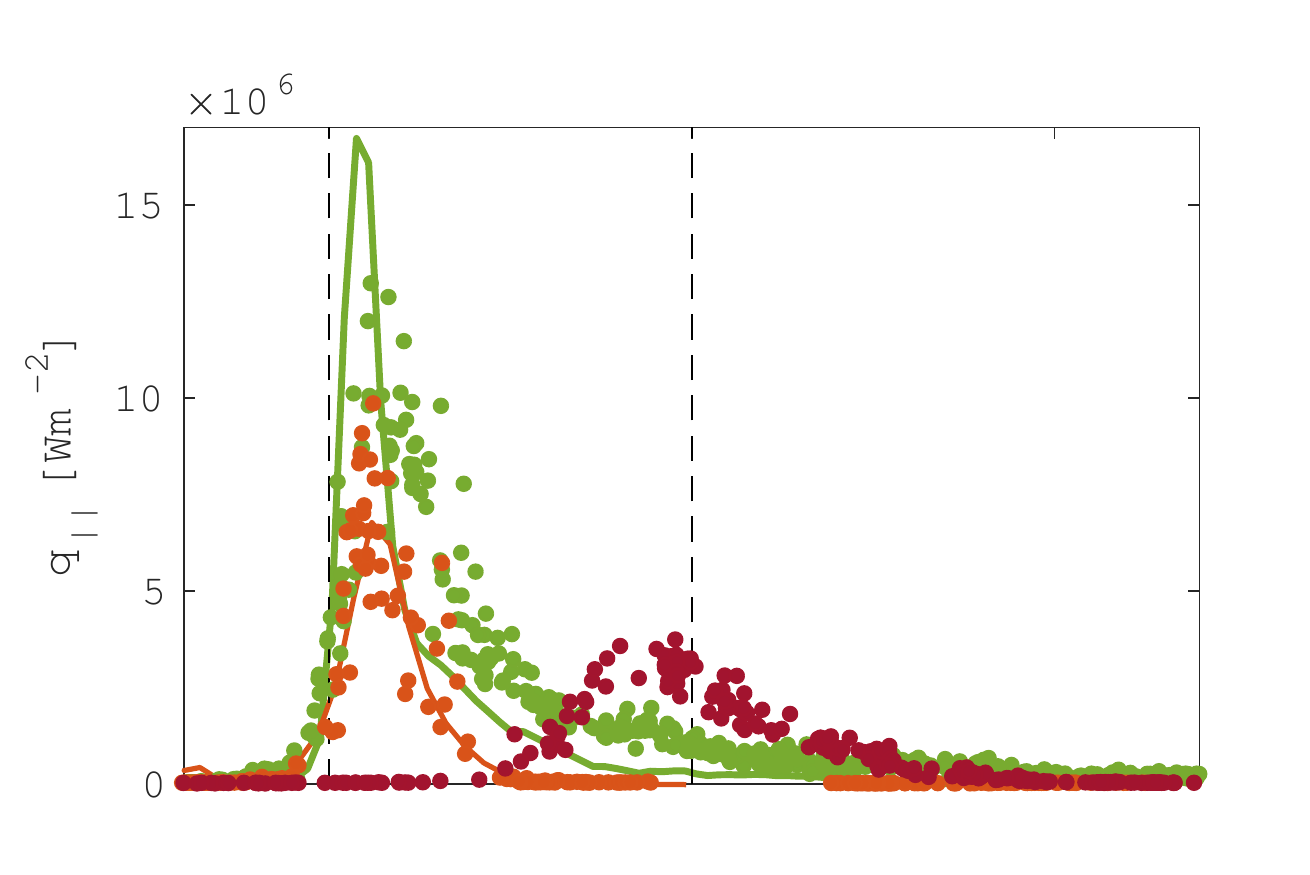}
			\put(37,25){\color{matlab_orange}SP2 (SF-LFS)}
			\put(55,13){\color{matlab_burgundy}SP4 (SF-LFS)}
			\put(33,35){\color{matlab_green}OSP (SN)}
			\put(82,42){\color{black}(b)}
			\put(93,4){\rotatebox{90}{\color{matlab_green}\tiny \#70320, \textbf{\#70322}}}
		\end{overpic}
	\end{subfigure}
	\begin{subfigure}{0.5\textwidth}
		\centering
		\vspace{-0.15cm}
		\begin{overpic}[width=8.8cm]{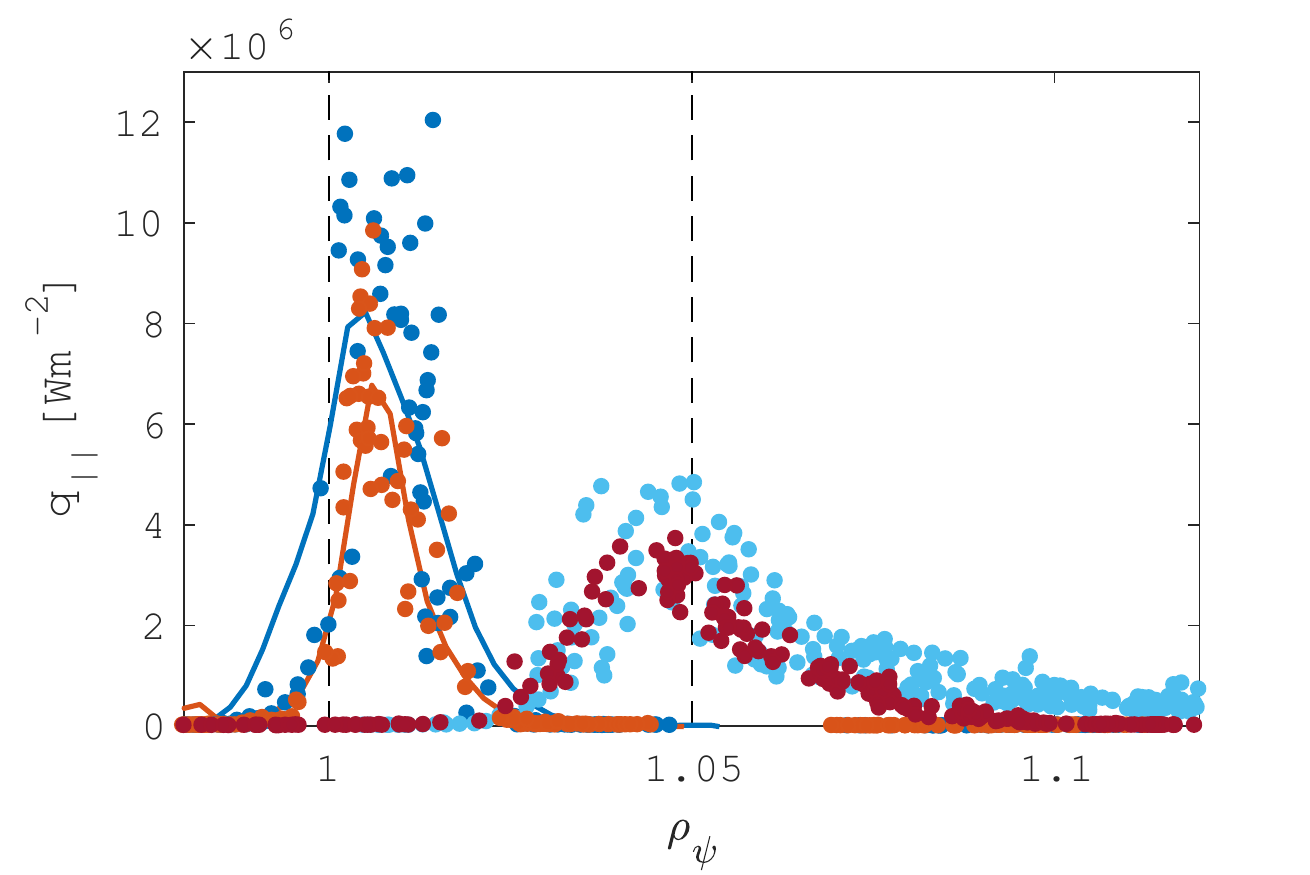}
			\put(35,34){\color{matlab_orange}SP2 (Baffled)}
			\put(60,14){\color{matlab_burgundy}SP4 (Baffled)}
			\put(35,39){\color{matlab_blue}SP2 (Unbaffled)}
			\put(60,19){\color{matlab_lightblue}SP4 (Unbaffled)}
			\put(82,42){\color{black}(c)}
			\put(93,4){\rotatebox{90}{\color{matlab_orange}\tiny \#70317, \#70319, \textbf{\#70497},}}
			\put(93,36){\rotatebox{90}{\color{matlab_blue}\tiny \textbf{\#67717}}}
		\end{overpic}
	\end{subfigure}
	\caption{Outer target heat flux profiles of the active outer strikepoints of the following geometries prior to any N$_2$ injection, measured by the LPs (markers) and the IR (solid lines): (a) the baffle-optimised SF-LFS ($dr_{\textrm{\scriptsize X2}}\sim 3.7$ mm) compared with the {baffled} SN; (b) the baffle-optimised SF-LFS ($dr_{\textrm{\scriptsize X2}}\sim 6.9$ mm) compared with the {baffled} SN; (c) the baffle-optimised SF-LFS ($dr_{\textrm{\scriptsize X2}}\sim 6.9$ mm) compared with the unbaffled counterpart. The dashed lines represent the positions of the primary and secondary separatrices of the SF-LFS.}
	\label{fig:LPprofs_peakheat}
\end{figure}

Compared with the baffled SN, a reduction in $q_{||}^{\textrm{\scriptsize peak}}$ is observed at the outer divertor in the baffle-optimised SF-LFS (see figure \ref{fig:LPprofs_peakheat} (a)-(b)), as seen in previous studies without baffles in TCV \cite{Maurizio2019e} {and in NSTX \cite{Soukhanovskii2012}}. Note that while the LPs and IR show fair agreement for measured $q_{||}$ profiles, discrepancies remain in the magnitude of $q_{||}^{\textrm{\scriptsize peak}}$, with the VIR reporting generally higher and the HIR generally lower values than the LP-measured $q_{||}^{\textrm{\scriptsize peak}}$.{  Despite the difference in magnitude, the LPs and IR show consistent trends in target heat flux with varying divertor geometry.} The SF-LFS case with $dr_{\textrm{\scriptsize X2}}\sim3.7$ mm presents a significant parallel heat flux reduction in the outer divertor with respect to the SN configuration, with IR measuring a reduction of $66\%$ and LPs of $57\%$. For the SF-LFS case with larger X-point separation ($dr_{\textrm{\scriptsize X2}}\sim6.9$ mm), IR sees a reduction of $59\%$, and LPs of $24\%$. This heat flux reduction coincides with a reduction in the $\Gamma_{||}^{\textrm{\scriptsize peak}}$ and $V_{\textrm{\scriptsize sh}}$, rather than a strong reduction in $T_e^{\textrm{\scriptsize peak}}$. Furthermore, this reduction with respect to the SN is stronger for the SF-LFS with small $dr_{\textrm{\scriptsize X2}}$  since the heat fluxes at SP2 and SP4 are more balanced. For the remainder of this paper, the focus will be placed upon the case with $dr_{\textrm{\scriptsize X2}}\sim6.9$ mm, while the small $dr_{\textrm{\scriptsize X2}}$ case will be used to investigate the generality of results to the SF-LFS.

When compared with the unbaffled configuration, the baffle-optimised SF-LFS presents a $q_{||}^{\textrm{\scriptsize peak}}$ reduction of $\sim18\%$ at SP2 and $\sim23\%$ at SP4, as shown in figure \ref{fig:LPprofs_peakheat}(c). LP data suggest that this reduction coincides strongly with a reduction in $T_e^{\textrm{\scriptsize peak}}${, as seen for increasing divertor closure in JT60U and Alcator C-mod \cite{Asakura1999,Lipschultz2007}}, with no strong change in $\Gamma_{||}^{\textrm{\scriptsize peak}}$. This is also seen for the SF-LFS case with small $dr_{\textrm{\scriptsize X2}}$ (not shown here), and is therefore assumed to be a general feature of the SF-LFS configuration. Over the range $dr_{\textrm{\scriptsize X2}}=[2.0,7.1]~$mm, the baffles reduce $q_{||}^{\textrm{\scriptsize peak}}$ almost symmetrically at each active outer strike-point, so the dependence of outer target $q_{||}^{\textrm{\scriptsize peak}}$ distribution on  $dr_{\textrm{\scriptsize X2}}$ is largely unaffected by baffles.

At the inner target, we note that all configurations exhibit comparable $q_{||}$ profiles, demonstrating that the outer target $q_{||}$ is not simply reduced from the change in in-out power sharing with changing geometry. We conclude that the baffled SF-LFS exhibits a clear peak outer target heat flux reduction, with respect to both the unbaffled SF-LFS and the baffled SN in the absence of N$_2$ injection.

\section{Evolution of divertor conditions with N$_2$ seeding}\label{sec:div_N2}

\hspace*{0.3cm}

The mitigation of target heat flux requires significant radiative losses in the divertor, which can be increased by impurity seeding. In this section, the effect of N$_2$ injection on the target heat fluxes and divertor radiated power is reported. The time-integrated N$_2$ flux is used as a proxy for divertor N$_2$ content, since the absolute divertor N$_2$ concentration cannot yet be measured on TCV. This quantity represents an upper limit for the divertor N$_2$ content, as it accounts for neither wall-retention, exhaust to the TCV pumps nor core penetration of the impurity, which cause the actual divertor N$_2$ content to be lower. Nevertheless, it is considered a useful parameter in comparing the detachment properties of the different configurations. In these scenarios, only partial detachment \cite{Kallenbach2015} can be achieved, as N$_2$ causes plasma disruption before pronounced detachment. The lowest $T_e^{\textrm{\scriptsize peak}}$ measured is 6.2 eV, similar to the minimum target temperatures measured in other N$_2$ seeding experiments on TCV \cite{Fevrier2020}.

\subsection{Target heat flux mitigation}

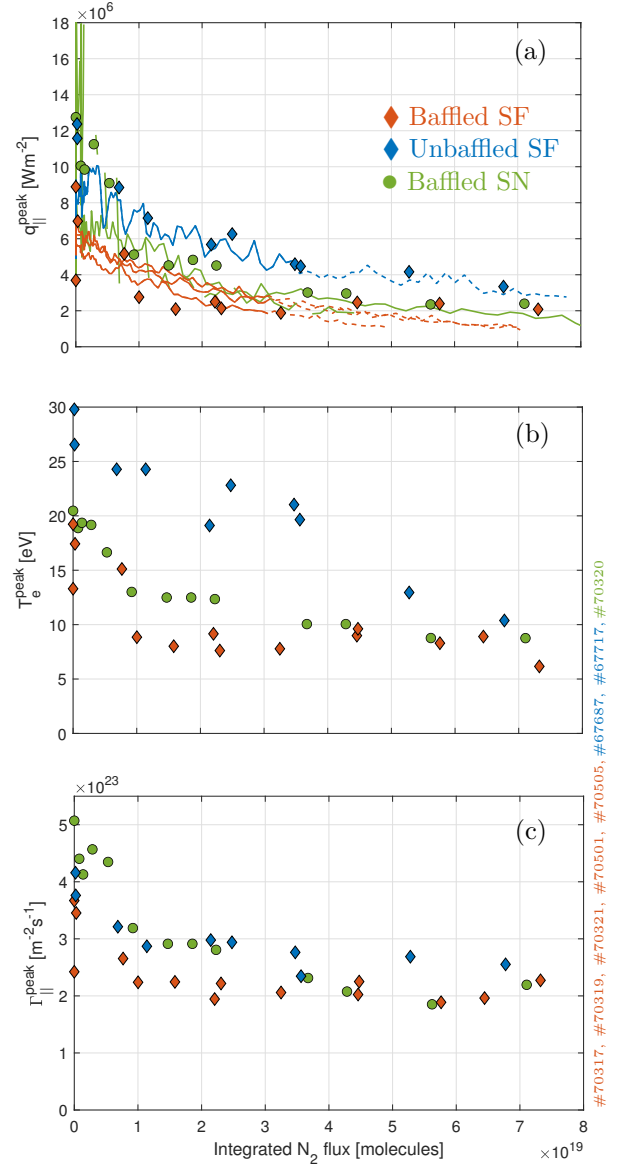
\begin{figure}[t]
	\centering
	\begin{subfigure}{0.5\textwidth}
		\centering
		\begin{overpic}[scale=0.6,percent]{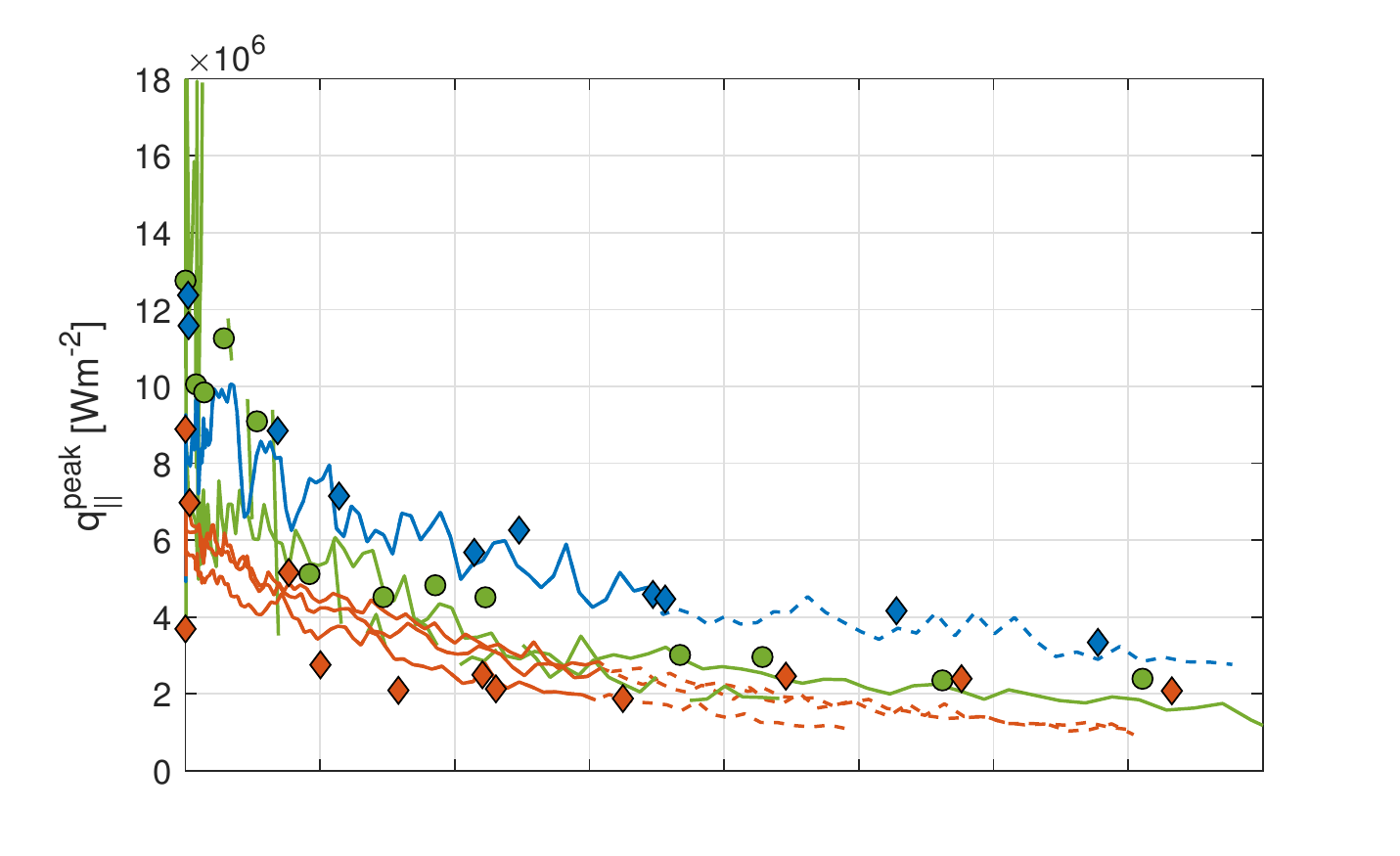}
			\put(80,50){\color{black}(a)}
			\put(60,35){\color{matlab_blue}$\blacklozenge$ Unbaffled SF}
			\put(60,40){\color{matlab_orange}$\blacklozenge$ Baffled SF}
			\put(60,30){\color{matlab_green}$\Circle[f]$ Baffled SN}
		\end{overpic}
	\end{subfigure}
	\begin{subfigure}{0.5\textwidth}
		\centering
		\begin{overpic}[scale=0.6,percent]{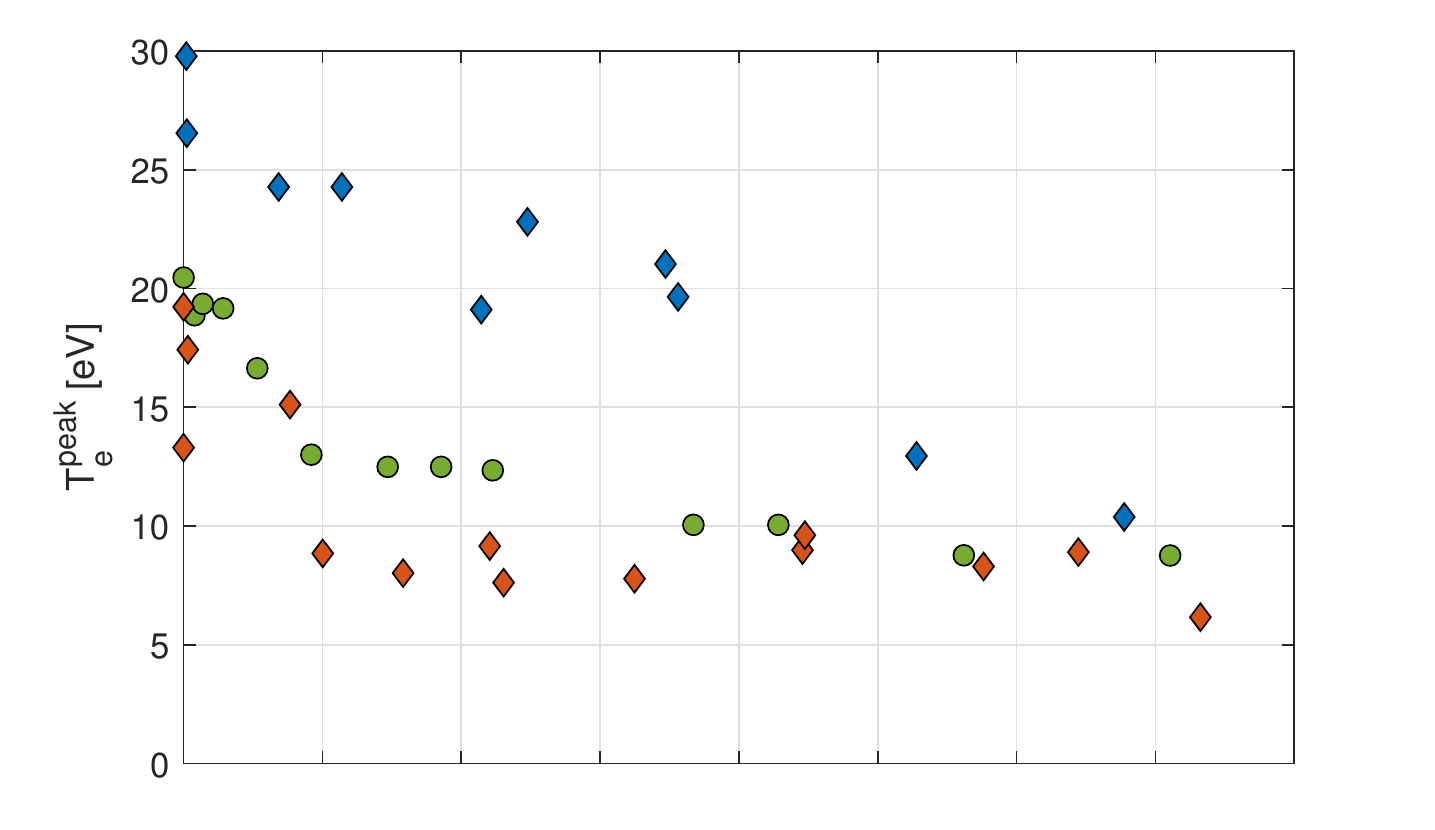}
			\put(80,49){\color{black}(b)}
			\put(92,1){\rotatebox{90}{\color{matlab_blue}\tiny \#67687, \#67717,}}
			\put(92,22){\rotatebox{90}{\color{matlab_green}\tiny \#70320}}
		\end{overpic}
	\end{subfigure}
	\begin{subfigure}{0.5\textwidth}
		\centering
		\begin{overpic}[scale=0.6,percent]{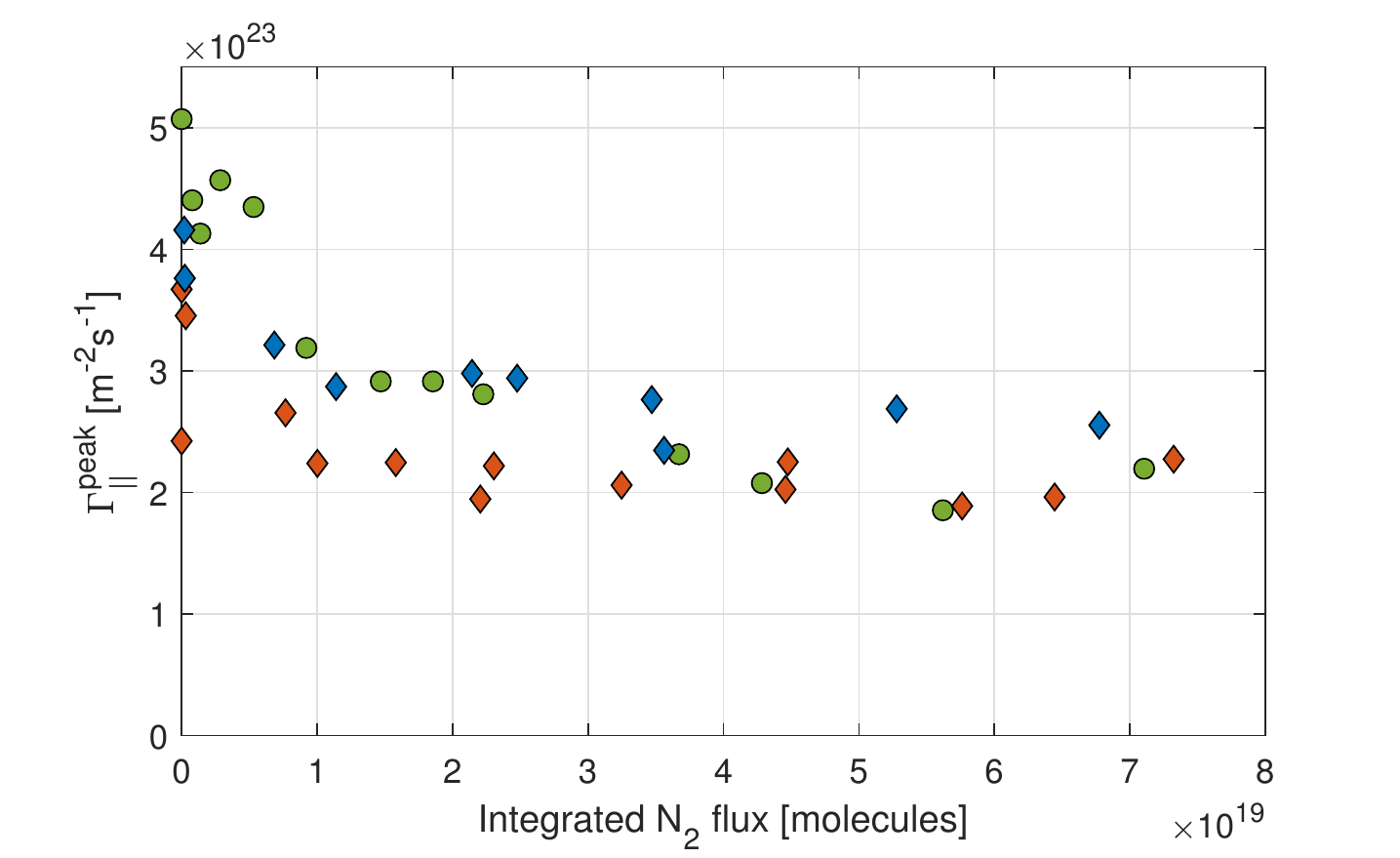}
			\put(80,50){\color{black}(c)}
			\put(92,9){\rotatebox{90}{\color{matlab_orange}\tiny \#70317, \#70319, \#70321, \#70501, \#70505,}}
		\end{overpic}
	\end{subfigure}
	\caption{Peak outer target (a) parallel heat flux, (b) electron temperature, (c) parallel ion flux, plotted as a function of the time-integrated N$_2$ flux. Data measured by LPs (markers) present the overall outer target peak value (considering both SP2 and SP4 for the SF-LFS), whereas heat flux data measured by IR (lines) consider only SP2, as SP4 data is not available. Dashed lines indicate where the outer target heat flux is expected to peak at SP4. Results are shown from repeat discharges in each geometry {to give an indication of experimental reproducibility}.}
	\label{fig:N2_targetevolution}
\end{figure}

In all configurations, a general decrease in peak target ion flux ($\Gamma_{||}^{\textrm{\scriptsize peak}}$), electron temperature ($T_e^{\textrm{\scriptsize peak}}$) and heat flux ($q_{||}^{\textrm{\scriptsize peak}}$) is observed in the outer divertor with N$_2$ injection as the outer divertors begin to detach (see figure \ref{fig:N2_targetevolution}). In the baffled SF-LFS, LP-measured $q_{||}^{\textrm{\scriptsize peak}}$ decreases strongly with the start of N$_2$ seeding, but quickly plateaus to an approximately constant value of $2.3\pm0.3$ MWm$^{-2}$. This represents the overall peak of the outer targets, which for the SF-LFS is the maximum of SP2 and SP4, and for the SN is simply the OSP peak. Peak target temperature and ion flux behave similarly, attaining approximately constant values of $8.2\pm1.0$ eV and $(2.1\pm 0.2)\times 10^{23}~\textrm{m}^{-2}\textrm{s}^{-1}$ respectively. Impurity seeding strongly reduces $q_{||}^{\textrm{\scriptsize peak}}$ at SP2, which is connected to the near SOL, but has a smaller effect at SP4, which is connected to the far SOL. For a time-integrated N$_2$ flow of approximately $3\times 10^{19}$ molecules, $q_{||}^{\textrm{\scriptsize peak}}$ at SP2 becomes lower than at SP4, while at SP4 $q_{||}^{\textrm{\scriptsize peak}}$ decreases very little with further N$_2$ injection. Divertor cooling with N$_2$ injection increases the neutral mean free path, potentially allowing neutrals to travel further into the confined plasma rather than remaining, and dissipating power, in the divertor. This may explain the observed $q_{||}^{\textrm{\scriptsize peak}}$ stagnation. Figure \ref{fig:N2_targetevolution} (a) also displays IR-measured $q_{||}^{\textrm{\scriptsize peak}}$, but only for SP2 as data is not available for SP4. From a time-integrated N$_2$ flow of $\sim3\times 10^{19}$ molecules, the IR data is marked by a dashed line for the SF-LFS configurations, as the SP2 peak is no longer expected to represent the overall peak of the outer targets. The IR data before this line is in fair agreement with the LPs, within the experimental scatter.

When compared with the unbaffled SF-LFS, the baffled SF-LFS exhibits a clear decrease in $q_{||}^{\textrm{\scriptsize peak}}$ {from} zero to moderate levels of N$_2$ seeding, coincidentially with a reduction in $T_e^{\textrm{\scriptsize peak}}$. For high injection levels of N$_2$, the difference in both $q_{||}^{\textrm{\scriptsize peak}}$ and $T_e^{\textrm{\scriptsize peak}}$ between the configurations is reduced, where the unbaffled SF-LFS displays continuous reductions, but the values remain constant in the baffled SF-LFS. The colder baffled divertor may reduce the radiative efficiency of N$_2$, hence stagnating the reduction of peak target heat flux. This will be discussed further in the following sub-section. However, the stagnation of electron temperature may also be due to a limitation in the LP measurement, where a modification of the plasma resistance or fluctuations in the floating potential may lead to an over-estimation of the electron temperature in detached conditions \cite{Ohno2001}.

Compared with the baffled SN, we see a clear reduction of $q_{||}^{\textrm{\scriptsize peak}}$ in the baffled SF-LFS until a time-integrated N$_2$ flux of $\sim4\times 10^{19}$ molecules. This heat flux reduction comes with only a slight reduction in $T_e^{\textrm{\scriptsize peak}}$ --- the presence of baffles appears to have a stronger effect on $T_e^{\textrm{\scriptsize peak}}$ than this change in magnetic geometry. Injecting additional N$_2$ tends to make $q_{||}^{\textrm{\scriptsize peak}}$ equal for the two configurations, {along with $\Gamma_{||}^{\textrm{\scriptsize peak}}$.} Note that the stagnation of $q_{||}^{\textrm{\scriptsize peak}}$ with N$_2$ seeding in the baffled SN occurs at a higher quantity of injected N$_2$ than for the baffled SF-LFS. {The reason for the equivalence of target heat flux between geometries with strong N$_2$ seeding is not currently understood. This could be an indication of lower divertor N$_2$ retention in the SF-LFS compared to the SN.}

At the inner strike-point, the peak target heat flux is comparable across all geometries, including those with N$_2$ injection. We conclude that the heat flux reduction observed in the baffled SF-LFS with respect to all other cases is not simply due to a redirection of power to the inner target.

\subsection{Divertor radiated power}

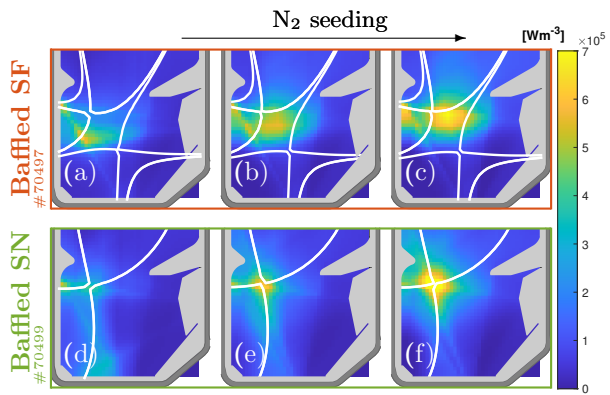
\begin{figure}[t]
	\centering
	\begin{overpic}[scale=0.45,percent]{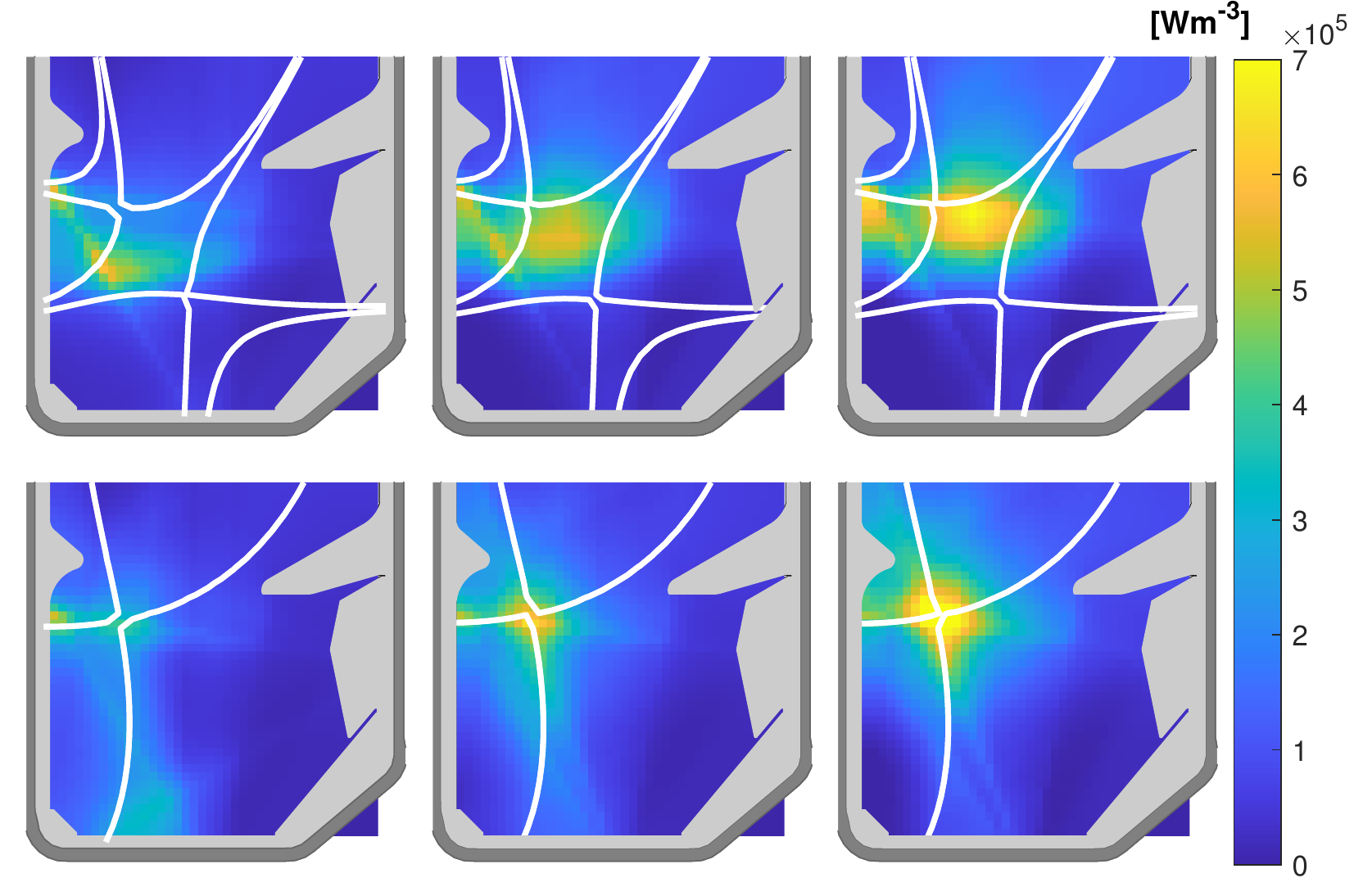}
		\put(4,39){\color{white}\textbf(a)}
		\put(34,39){\color{white}\textbf(b)}
		\put(64,39){\color{white}\textbf(c)}
		\put(2,62){\color{matlab_orange}\thicklines \line(2,0){88}}
		\put(2,34){\color{matlab_orange}\thicklines \line(2,0){88}}
		\put(2,33.85){\color{matlab_orange}\thicklines \line(0,2){28.35}}
		\put(90,33.85){\color{matlab_orange}\thicklines \line(0,2){28.35}}
		\put(4,7){\color{white}\textbf(d)}
		\put(34,7){\color{white}\textbf(e)}
		\put(64,7){\color{white}\textbf(f)}
		\put(2,30.5){\color{matlab_green}\thicklines \line(2,0){88}}
		\put(2,2.5){\color{matlab_green}\thicklines \line(2,0){88}}
		\put(2,2.3){\color{matlab_green}\thicklines \line(0,2){28.35}}
		\put(90,2.3){\color{matlab_green}\thicklines \line(0,2){28.35}}
		\put(25,64){\color{black}\vector(2,0){50}}
		\put(41,66){\color{black}\small N$_2$ seeding}
		\put(41,66){\color{black}\small N$_2$ seeding}
		\put(-1,3){\rotatebox{90}{\color{matlab_green}\tiny \#70499}}
		\put(-1,34){\rotatebox{90}{\color{matlab_orange}\tiny \#70497}}
		\put(-5,36){\rotatebox{90}{\color{matlab_orange}\textbf{Baffled SF}}}
		\put(-5,5){\rotatebox{90}{\color{matlab_green}\textbf{Baffled SN}}}
	\end{overpic}
	\caption{Poloidal plots of plasma emissivity from bolometry for (a)-(c): the baffled SF-LFS, and (d)-(f): the baffled SN configurations. From left to right, the quantity of time-integrated N$_2$ flux is increased: (a),(d): $\sim0$ molecules; (b),(e): $\sim3.0\times 10^{19}$ molecules; (c), (f): $\sim6.0 \times 10^{19}$ molecules.}
	\label{fig:emiss_2D}
\end{figure}

The plasma emissivity calculated from tomographic inversions \cite{Kamleitner2015} of the bolometry data gives an indication of regions of strong radiated power, as shown in figure \ref{fig:emiss_2D}. The baffled SF-LFS exhibits a radiation region between the two X-points, whereas an X-point radiator is seen for the baffled SN configuration. The inter-null radiation region is observed in the SF-LFS for both X-point separations, with and without baffles, and before and during the N$_2$-seeded phase of the discharge (the small $dr_\textrm{\scriptsize X2}$ and unbaffled cases are not shown here). {This broadly distributed radiation region in the SF-LFS is also observed in previous TCV experiments \cite{Reimerdes2017a}, as well as in other SF configurations and in other tokamaks \cite{Soukhanovskii2012,Soukhanovskii2018a,Soukhanovskii2022}.} This is therefore a key feature of the SF geometry. These regions increase in size and intensity with N$_2$ seeding, moving closer to the confined plasma region, as apparent in figure \ref{fig:emiss_2D}.

\begin{figure}[t]
	\centering
	\hspace{-0.5cm}
	\begin{overpic}[scale=0.6,percent]{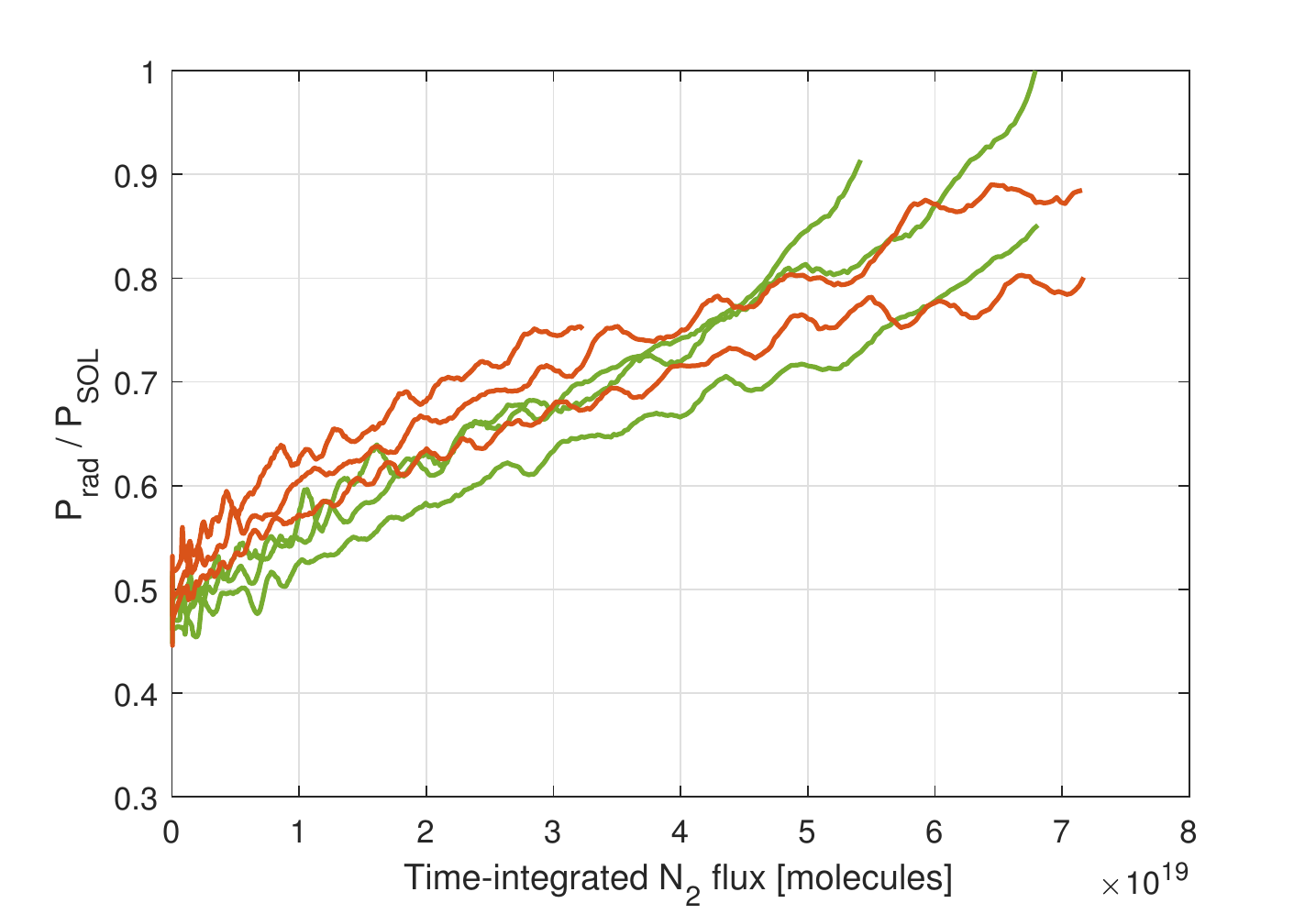}
		\put(15,40){\color{matlab_orange}Baffled SF}
		\put(30,25){\color{matlab_green}Baffled SN}
		\put(92,9){\rotatebox{90}{\color{matlab_orange}\tiny \textbf{\#70321}, \#70497, \#70501}}
		\put(95,9){\rotatebox{90}{\color{matlab_green}\tiny \textbf{\#70322}, \#70499, \#70504}}
		
		\put(64,9){\includegraphics[scale=.2]{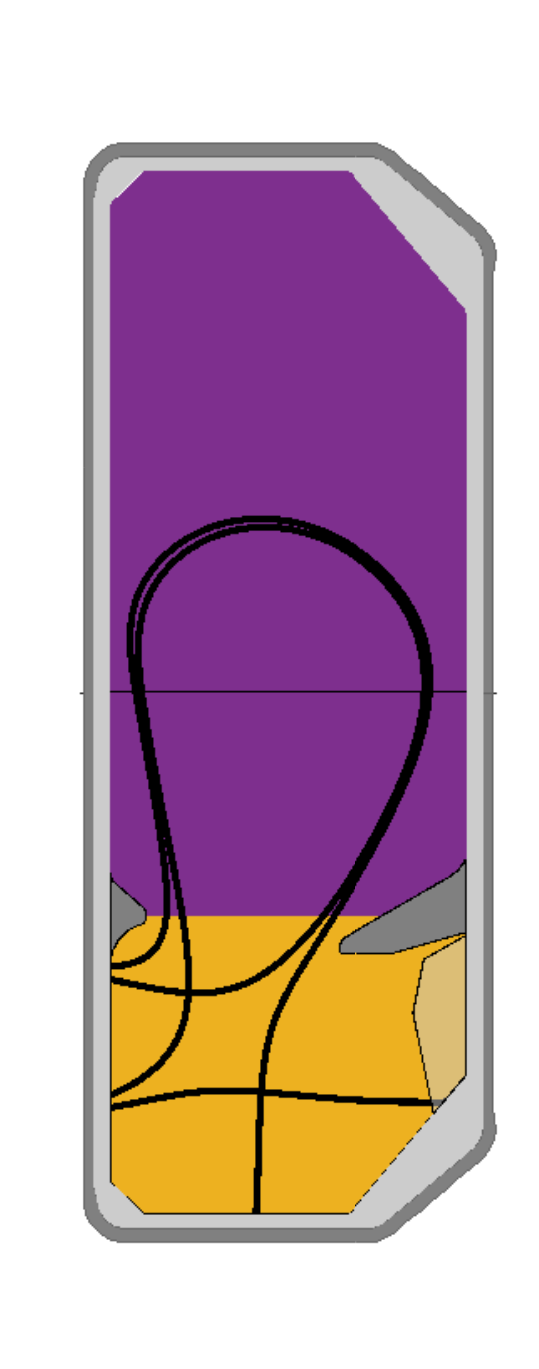}}
		\put(76,11){\color{matlab_orange}\thicklines \line(0,2){27}}
		\put(65.5,38){\color{matlab_orange}\thicklines \line(2,0){10.5}}
		\put(65.5,11){\color{matlab_orange}\thicklines \line(2,0){10.5}}
		\put(65.5,11){\color{matlab_orange}\thicklines \line(0,2){27}}
				
		\put(76.5,9){\includegraphics[scale=.2]{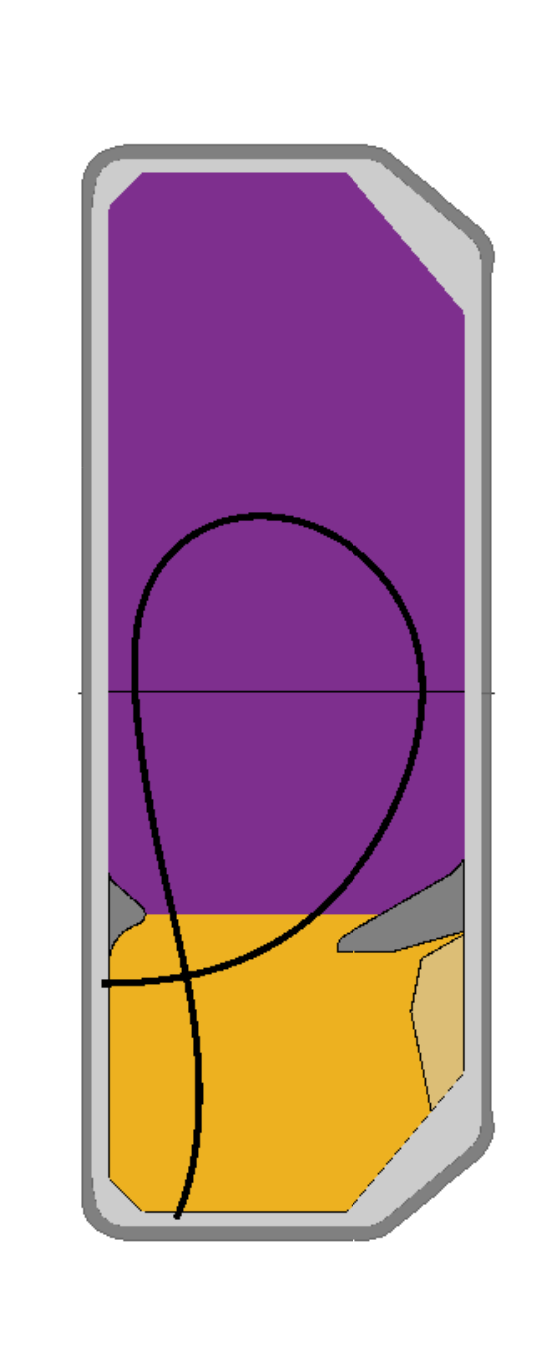}}
		\put(88.5,11){\color{matlab_green}\thicklines \line(0,2){27}}
		\put(78,38){\color{matlab_green}\thicklines \line(2,0){10.5}}
		\put(78,11){\color{matlab_green}\thicklines \line(2,0){10.5}}
		\put(78,11){\color{matlab_green}\thicklines \line(0,2){27}}
	\end{overpic}
	\caption{Bolometry data from the recently-installed RADCAM system showing the proportion of the power entering the SOL, $P_{\textrm{\scriptsize SOL}}$, that is radiated in the divertor and X-point region for the baffled SF-LFS and SN configurations. Results are shown from repeat discharges in each geometry. Inset, for each magnetic geometry, the regions corresponding to the core (purple) and divertor/X-point region (yellow) for the purpose of the radiated power fraction calculation are shown.}
	\label{fig:N2_prad_psol}
\end{figure}
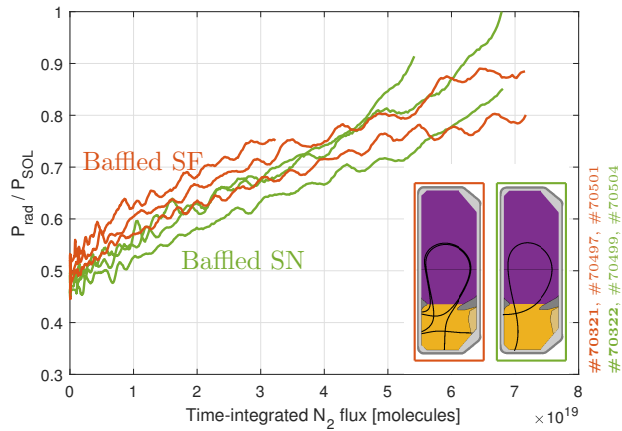

The radiated exhaust power, $P_{\textrm{\scriptsize rad}}$, is calculated by integrating the plasma emissivity over the divertor volume and X-point region (see yellow region inset in figure \ref{fig:N2_prad_psol}). To account for small variations in the Ohmic heating power and core conditions in each configuration, we compare the radiated power fraction, by evaluating the ratio of $P_{\textrm{\scriptsize rad}}$ to the power entering the SOL, $P_{\textrm{\scriptsize SOL}}$. $P_{\textrm{\scriptsize SOL}}$ is the input (Ohmic) heating power minus the core radiated power. For simplicity, the core is defined as the region above the HFS baffle tip, as shown by the purple region inset in figure \ref{fig:N2_prad_psol}, excluding the X-point region while fairly comparing two magnetic geometries with different core shapes. The divertor radiated power fraction thus obtained (figure \ref{fig:N2_prad_psol}) is at most $10\%$ higher in the baffled SF-LFS than in the baffled SN configuration, up to a time-integrated N$_2$ flux of approximately $4\times 10^{19}$ molecules. {This is consistent with the 10-15\% difference in divertor radiated power measured in DIII-D \cite{Soukhanovskii2015}.} Similar values are obtained for the small $dr_{\textrm{\scriptsize X2}}$ case (not shown) within the experimental scatter. {However, the uncertainty associated with the divertor bolometric inversions for these discharges is around $10\%$, and so no strong difference in $P_{\textrm{\scriptsize rad}}$ can be concluded between the two configurations.}

\section{Core-divertor compatibility}\label{sec:core_conf}

\hspace*{0.3cm}

As part of the assessment of the SF-LFS configuration as a potential divertor solution for future reactors, we must consider divertor-core compatability alongside power exhaust enhancements, to limit core impurity pollution and the degradation of energy confinement. Two complementary metrics will be used in this assessment: the core effective charge and the main plasma energy confinement time. 

\subsection{Core effective charge}

The core effective charge is defined in the usual manner as,
\begin{equation}
Z_{\textrm{\scriptsize eff}}= \frac{\Sigma_i n_i Z_i^2}{n_e},
\end{equation}
where $n_i$ is the density of each ion species $i$ and $Z_i$ is the charge number of that species. $Z_{\textrm{\scriptsize eff}}$ is estimated from measurements of $V_{\textrm{\scriptsize loop}}$ and the profile-averaged core temperature from Thomson scattering, assuming steady state conditions and neo-classical conductivity \cite{Sauter1999,Sauter2001}. 

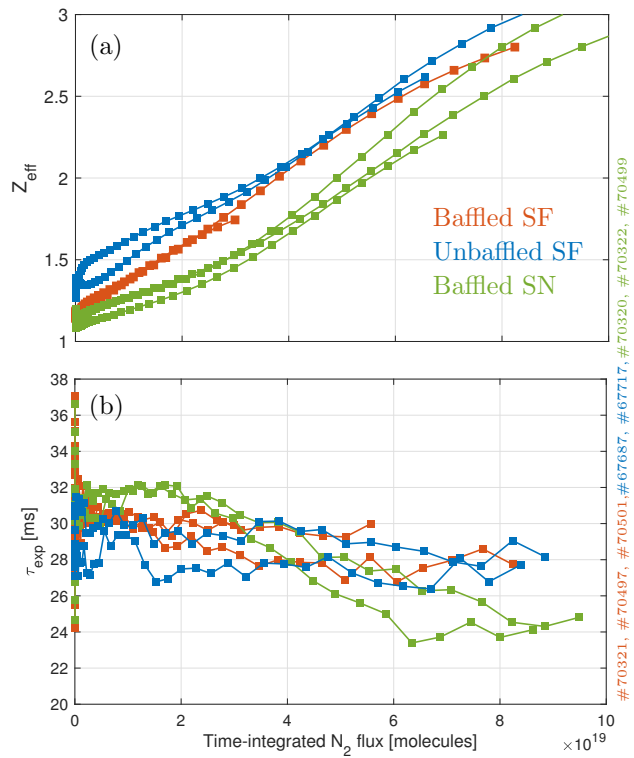
\begin{figure}[t]
	\centering
	\begin{subfigure}{0.5\textwidth}
		\centering
		\begin{overpic}[scale=0.6,percent]{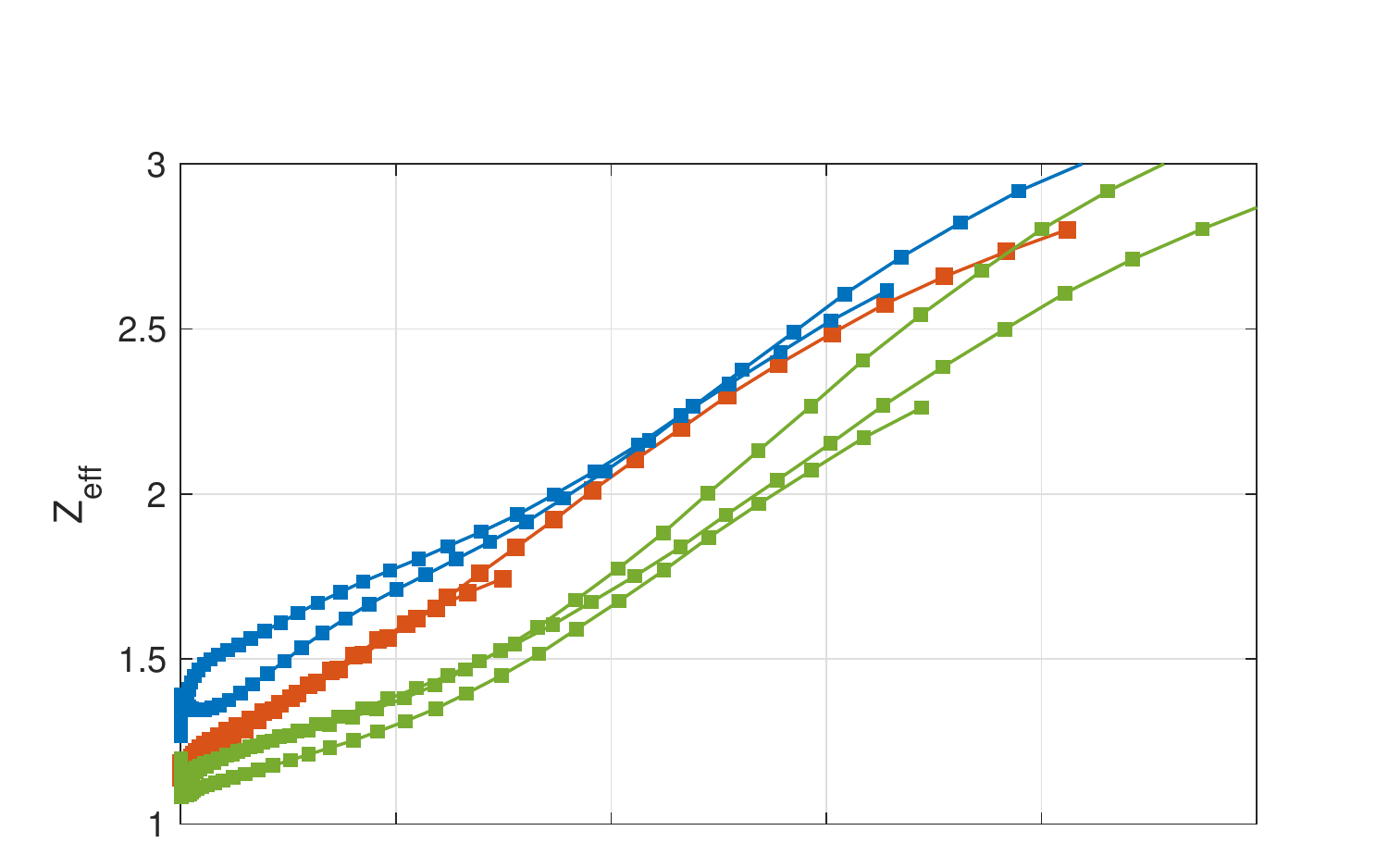}
			\put(15,45){\color{black}(a)}
			\put(65,15){\color{matlab_blue}Unbaffled SF}
			\put(65,20){\color{matlab_orange}Baffled SF}
			\put(65,10){\color{matlab_green}Baffled SN}
		\end{overpic}
	\end{subfigure}
	\begin{subfigure}{0.5\textwidth}
		\centering
		\begin{overpic}[scale=0.6,percent]{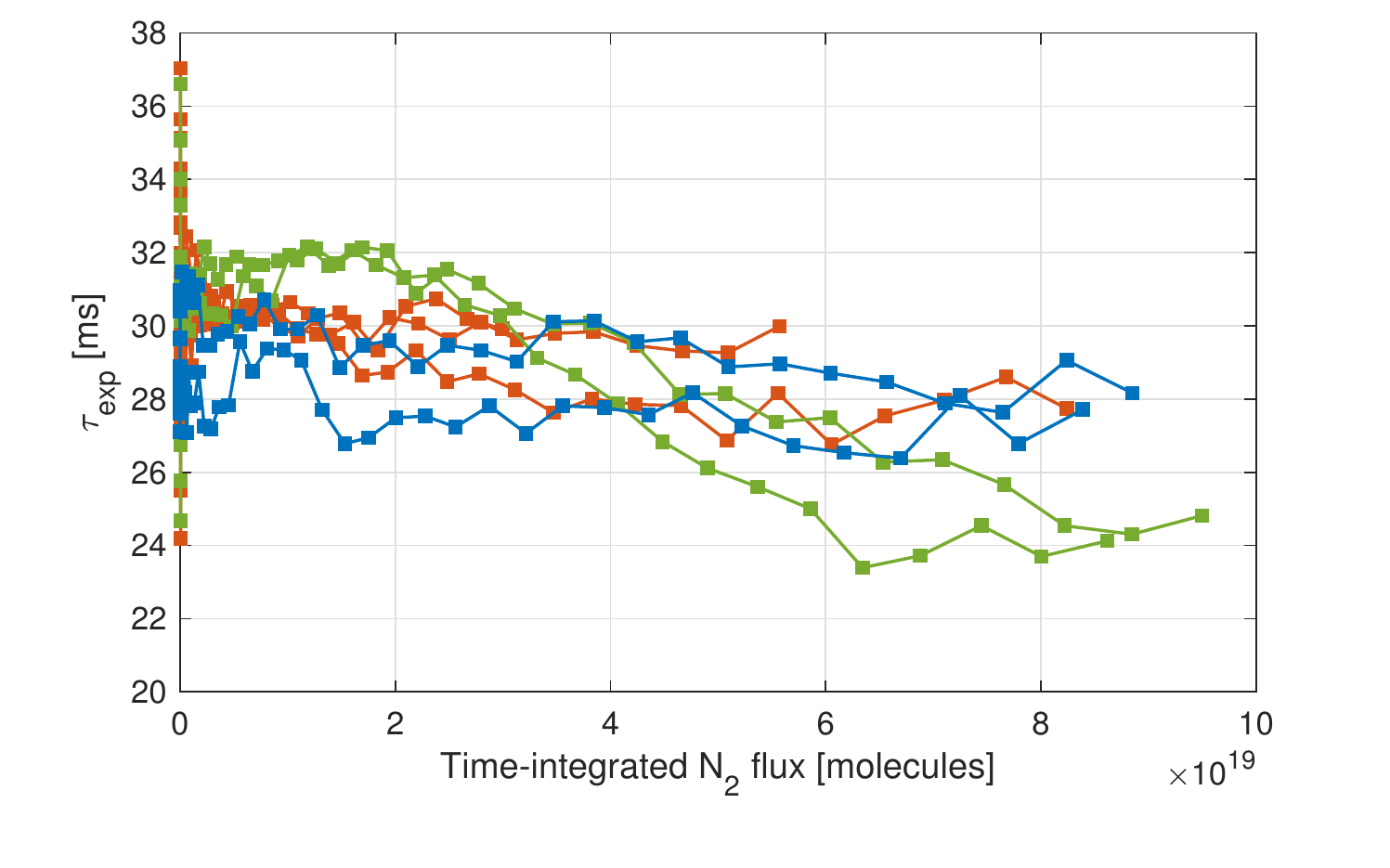}
			\put(15,55){\color{black}(b)}
			\put(92,13){\rotatebox{90}{\color{matlab_orange}\tiny \#70321, \#70497, \#70501,}}
			\put(92,43){\rotatebox{90}{\color{matlab_blue}\tiny \#67687, \#67717,}}
			\put(92,63){\rotatebox{90}{\color{matlab_green}\tiny \#70320, \#70322, \#70499}}
		\end{overpic}
	\end{subfigure}
	\caption{(a) Core effective charge, $Z_{\textrm{\scriptsize eff}}$, estimated from measurements of $V_{\textrm{\scriptsize loop}}$ and the profile-averaged core temperature from Thomson scattering. (b) Core energy confinement time estimated from Thomson scattering measurements of core plasma density and temperature profiles. Results are shown from repeat discharges in each geometry. }
	\label{fig:core_div}
\end{figure}

The core effective charge is shown in figure \ref{fig:core_div} (a), as a function of the time-integrated N$_2$ flux. Initially, $Z_{\textrm{\scriptsize eff}}$ is only slightly above 1, the value of an undiluted deuterium plasma, but then increases once N$_2$ seeding begins. Measurements of core carbon density suggest that this increase in $Z_{\textrm{\scriptsize eff}}$ is in fact due to an increase in core nitrogen content, rather than a variation in electron or carbon density. 

In the SF-LFS configuration, baffles only slightly reduce $Z_{\textrm{\scriptsize eff}}$. We expect that while the colder divertor would be more transparent to impurities, the increased divertor closure would simply decrease the probability of impurities entering the core plasma. Comparing with the baffled SN, the baffled SF-LFS configuration exhibits an up to $25\%$ higher $Z_{\textrm{\scriptsize eff}}$, suggesting stronger core impurity penetration. The SF-LFS case with small $dr_{\textrm{\scriptsize X2}}$ {(not shown here)} displays a slightly milder trend, but still with an up to $20\%$ increase in $Z_{\textrm{\scriptsize eff}}$. {In Ne-seeded near-exact SF experiments on TCV, no difference was discerned in the core impurity shielding of the SF and SN \cite{Reimerdes2015b}. SF- experiments in NSTX observe a reduction in core impurity content during the formation of the SF configuration, but this is attributed to a difference in ELM behaviour rather than the magnetic geometry \cite{Soukhanovskii2011,Soukhanovskii2016}. }

{The increase in Z$_{\textrm{\scriptsize eff}}$ observed here} may signify a SOL in the {baffled L-mode} SF-LFS that is more transparent to impurities than for the baffled SN. Furthermore, this suggests that a radiation region located further away from the core plasma, figure \ref{fig:emiss_2D}, did not necessarily improve core compatibility. This will need to be explored in future experiments under higher power conditions, where low-Z impurity screening can increase \cite{Mccracken1997}. 

\subsection{Energy confinement time}

We now focus on the energy confinement time, $\tau_{\textrm{\scriptsize exp}}$, defined as,
\begin{equation}
\tau_{\textrm{\scriptsize exp}} = \frac{W_{\textrm{\scriptsize MHD}}}{P_{\textrm{\scriptsize in}}},
\end{equation}
where $W_{\textrm{\scriptsize MHD}}$ is the plasma stored energy and $P_{\textrm{\scriptsize in}}$ the input heating power, which in this case is only Ohmic. Although diamagnetic loop measurements are often used to measure the stored plasma energy, it is difficult to compensate the flux measurement sensitivity to poloidal magnetic fields \cite{Manini2002}, which, across the studied configurations, can differ greatly. Here, $W_{\textrm{\scriptsize MHD}}$ is estimated by integrating the core electron temperature and density profiles, from Thomson scattering, over the plasma volume,
\begin{equation}
W_{\textrm{\scriptsize MHD}}=2\pi \iint(3n_eT_e) R dR dZ,
\end{equation}
where $R$ and $Z$ are the poloidal coordinates. We assume $T_i=T_e$ and that impurity densities are in the trace limit. The latter assumption can lead to an overestimation of $W_{\textrm{\scriptsize MHD}}$ by up to $15\%$. CXRS (Charge eXchange Recombination Spectroscopy) measurements of $T_i$ were only available for a small subset of the data presented, and so are not used, but confirm the trends observed by the assumption $T_i=T_e$. 

Figure \ref{fig:core_div} (b) shows the energy confinement time for the discharges presented in the study. Initially, the baffled SF-LFS shows a slightly lower energy confinement time than the baffled SN ($11\%$ at most). As {further} N$_2$ is injected, this difference reverses, with the SF-LFS now showing a $\sim16\%$ increase in $\tau_{\textrm{\scriptsize exp}}$ compared to the SN. {Snowflake experiments in DIII-D, NSTX and MAST-U find that core confinement is unaffected by the SF geometry in attached conditions \cite{Soukhanovskii2012,Soukhanovskii2022, Soukhanovskii2016}. Meanwhile, radiative experiments in DIII-D observe a stronger degradation of stored energy in the SF compared to the SN \cite{Soukhanovskii2018a,Hill2013}. 

The presence of baffles in the SF-LFS does not strongly affect the core confinement (consistent with SN experiments in JT60U  \cite{Asakura1999}). In TCV,} we conclude that no strong, systematic difference between the three configurations is discerned {in these low density, L-mode experiments. Analysis of SF-LFS discharges at higher plasma density and in H-mode is planned, where the difference in magnetic configuration is expected to have a more significant effect on core confinement.}

\section{Conclusion}\label{sec:conc}

\hspace*{0.3cm}

Baffled SF-LFS geometries have been successfully developed on TCV to explore the effect of divertor closure on the power exhaust and divertor-core compatibility of the SF-LFS, compared with standard baffled SN configurations. In the Ohmic L-mode discharges investigated herein, the SF-LFS shows a peak outer target heat flux reduction of around $20\%$ with baffling, and maintains the inter-null radiation region observed without baffles, with no significant effect on core-divertor compatibility. {In these proof-of-principle experiments, t}he baffled SF-LFS presents promising power exhaust features over the baffled SN in terms of outer target heat flux reductions and balance of strike-point heat flux distribution.  However, whilst approaching detachment through injecting N$_2$, these differences diminish, with similar target conditions and divertor radiative power losses between the two geometries for the higher N$_2$ injection rates. Despite a radiation region located further from the confined plasma for the baffled SF-LFS, no meaningful improvement is observed in either core confinement or core effective charge as a function of time-integrated N$_2$ flux.

{The experimental scenarios presented herein form a good reference for detailed model validations and extrapolations, exploring important physics that has not yet been understood from the experimental results alone. Moving forward, ongoing EMC3-EIRENE simulations \cite{Lunt2016b} look to further explore the roles of divertor radiative losses and core impurity shielding in target heat flux mitigation, whilst studying the dependence of divertor cross-field transport on magnetic geometry. In parallel, further experiments will explore the effect of increasing SOL opacity to neutrals, by increasing the separatrix density and input power, to investigate whether the SF-LFS benefits can be maintained while seeding impurities.}

\ack 
This work was supported in part by the Swiss National Science Foundation. This work has been carried out within the framework of the EUROfusion Consortium, funded by the European Union via the Euratom Research and Training Programme (Grant Agreement No 101052200 — EUROfusion). Views and opinions expressed are however those of the author(s) only and do not necessarily reflect those of the European Union or the European Commission. Neither the European Union nor the European Commission can be held responsible for them. 

\section*{Data availability statement}
The data used to produce the figures in this article are available at \href{https://zenodo.org/record/6957736}{https://doi.org/10.5281/zenodo.6957736}.

\section*{References}
\bibliographystyle{iopart-num}
\bibliography{references}

\end{document}